\documentclass[preprint2]{proto}
\usepackage{times}
\usepackage{natbib}
\bibpunct{(}{)}{;}{a}{,}{,}
\newcommand{\refs}{\par\noindent\hangindent=1pc\hangafter=1}
\def\msun{M_\odot}
\def\mjup{M_{\rm Jup}}
\def\rjup{R_{\rm Jup}}

\def\mearth{\,{\rm M}_\oplus}
\def\rearth{\,{\rm R}_\oplus}

\def\te{T_{\rm eff}}

\def\simgr{\,\hbox{\hbox{$ > $}\kern -0.8em \lower 1.0ex\hbox{$\sim$}}\,}
\def\simle{\,\hbox{\hbox{$ < $}\kern -0.8em \lower 1.0ex\hbox{$\sim$}}\,}
\def\gcc{\,\,\,{\rm g}\,{\rm cm}^{-3}}             
\def\gc2{\,\,\,{\rm g}\,{\rm cm}^{-2}}             
\newcommand{\me}{$M_{\oplus}$}

\newcommand{\ttherm}{$T_{\rm therm}$}
\newcommand{\teq}{$T_{\rm eq}$}

\newcommand{\tint}{$T_{\rm int}$}

\newcommand{\cp}{\citep}
\newcommand{\ct}{\citet}

\begin{document}

\title{\textbf{\LARGE Planetary internal structures}}

\author {\textbf{\large Isabelle Baraffe}}
\affil{\small\em University of Exeter, Physics and Astronomy, Exeter, United Kingdom}
\author {\textbf{\large Gilles Chabrier}}
\affil{\small\em Ecole Normale Sup\'erieure de Lyon, CRAL, France \and
University of Exeter, Physics and Astronomy, Exeter, United Kingdom}
\author {\textbf{\large Jonathan Fortney}}
\affil{\small\em University of California, Astronomy and Astrophysics, Santa Cruz, United States of America}
\author {\textbf{\large Christophe Sotin}}
\affil{\small\em JPL, Caltech, Pasadena, United States of America}

\begin{abstract}
\baselineskip = 11pt
\leftskip = 0.65in 
\rightskip = 0.65in
\parindent=1pc
{\small 
This chapter reviews the most recent advancements on the topic of terrestrial and giant planet interiors, including Solar System and extrasolar objects. Starting from an observed mass-radius diagram for known planets in the Universe, we will discuss the various types of planets appearing in this diagram and describe internal structures for each type. The review will summarize the status of theoretical and experimental works performed in the field of equation of states (EOS) for materials relevant to planetary
interiors and will address the main theoretical and experimental uncertainties and challenges. It will discuss the impact of new EOS on interior structures and bulk composition determination. We will discuss important dynamical processes which strongly impact the interior and evolutionary properties of planets (e.g plate tectonics, semiconvection) and describe non standard models recently suggested for our giant planets. We will address the case of short-period, strongly irradiated exoplanets
and critically analyse some of the physical mechanisms which have been suggested to explain their
anomalously large radius. 
 \\~\\~\\~}

\end{abstract}

\section{\textbf{INTRODUCTION}}

\begin{figure}[h]
\vspace{-1.5cm}
  \includegraphics[scale=0.35]{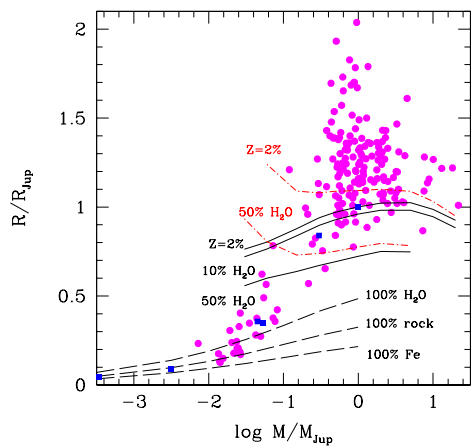}
  \vspace{-2.5cm}
 \caption{\small Mass versus radius of  known planets, including Solar System  planets (blue squares: Jupiter, Saturn, Neptune, Uranus, Earth and Mars) and transiting exoplanets (magenta dots, from the Extrasolar Planet Encyclopedia exoplanet.eu). Note that the sample that is shown in this figure has  been cleaned up and
only planets where the parameters are reasonably precise are included. The curves correspond to models from 0.1 $\mearth$ to 20 $\mjup$ at 4.5 Gyr with various internal chemical compositions. The solid curves correspond to a mixture of H, He and heavy elements \cp[models from][]{Baraffe08}. The long dashed lines correspond to models composed of pure water, rock or iron from \cite{Fortney07a}.
 The "rock" composition here is olivine  (forsterite Mg$_2$SiO$_4$) or dunite. Solid  and long-dashed lines (in black)
are for non-irradiated models. Dash-dotted (red) curves correspond to
irradiated models at 0.045 AU from a Sun.
}  
\label{fig_mr}
 \end{figure}

The nineties were marked by two historical discoveries: the first planetary system around a star other than the Sun, namely a pulsar  \citep{Wolszczan92} and the first Jupiter-mass companion to a solar-type star \citep{Mayor95}. Before that,
the development and application of planetary structure theory was restricted to the few planets belonging to our Solar System.
Planetary interiors provide natural laboratories to study materials under high pressure, complementing experiments which can be done on Earth.  
This explains why planets have long been of interest to physicists studying the equation of states of hydrogen, helium and other heavy materials made of water, silicates or iron. 
Planets also provide laboratories complementary to stellar interiors  to study physical processes common to both families of objects, like semiconvection, tidal dissipation, irradiation or ohmic dissipation. 
 For Solar System planets, space missions and in situ explorations  have provided a wealth of information on their atmospheric composition, on ground compositions for terrestrial planets, and on gravitational fields for giant planets which provide constraints on their interior density distribution. 
The theory of planetary structures has thus long been built on our knowledge of our own planet, the Earth,  and of our neighbours in the Solar System.  The diversity of planetary systems revealed by the discoveries of thousands of exoplanet candidates now tells us that the Solar System is not an universal template for planetary structures and system architectures. The information collected on exoplanets will never reach the level of accuracy and details obtained for the Solar System planets. Current measurements are mostly limited to gross physical properties including the mass, radius, orbital properties and sometime some information on the atmospheric composition. Nevertheless, the lack of refined information is compensated by the large number of detected exoplanets, completing the more precise but restricted knowledge provided by our Solar System planets. The general theory describing planetary structures needs now to broaden and to account for more physical processes to describe the diversity of exoplanet properties  and to understand some puzzles. This diversity is     
illustrated by the mass-radius relation of known planets displayed in Fig. \ref{fig_mr}. Naively, one would expect a connection between the mass and the radius of a planet. But exoplanets tell us that knowing a planet mass does not specify its size, and vice versa. Figure \ref{fig_mr} shows that the interpretation of transit radii must be taken with caution since the mass range corresponding to
a given radius can span up to two orders of magnitudes. Conversely, a planet of the mass of Neptune, for example,  could have a variety of sizes, depending on its bulk composition and the mass of its atmosphere. Even more troubling is the existence of ambiguous conclusions about bulk composition, as illustrated by the properties of the exoplanet GJ 1214b, with mass 6.5  $\mearth$ and radius 2.5 $\rearth$, and which could be explained by at least three very different sets of structures \citep{Rogers10}. The study of planetary structures is a giant construction game where progress is at the mercy of advances in both experiment and theory of materials at high pressure (see \S 2), improvement in the knowledge of Solar System planets paced by exploratory missions (see \S 3 and \S 4), accumulation of data for a wide variety of exoplanets and our creativity to fill the shortfall of accurate  data and to interpret  some amazing properties (see \S 5). This chapter will describe in detail the recent building pieces which elaborate current theory of planetary internal structures.   Descriptions of  how models for planets are built and of their basic equations and ingredients have already been described in detail in previous reviews and will not be repeated here \cp[see][]{Guillot99, Fortney10, FortneyBaraffe11}.

\bigskip

\centerline{\textbf{ 2. EQUATIONS OF STATE OF PLANETARY}}

{\textbf{MATERIALS}}
\bigskip

The correct determination of the interior structure and evolution of  planets depends on 
the accuracy of the description of the thermodynamic properties of matter under the relevant conditions of temperature and
pressure. These latter reach up to about 20000 K and 70 Mbar (7000 GPa) for Jupiter typical central conditions.
While terrestrial planets (or Earth-like planets) are essentially composed of a solid/liquid core of heavy material with a thin atmosphere (see \S 3), giant planets  have  an envelope essentially composed of hydrogen and helium, with some heavier material enrichment, and a core of heavy elements (see \S 4). 
As a new term appearing with the discoveries of exoplanets, super-Earths  refer to objects with masses greater than the Earth's, regardless of their composition.
The heavier elements
consist of C, N and O, often referred to as "ices" under their molecule-bearing volatile forms (H$_2$O, the most abundant of these elements for
solar C/O and N/O ratios, CH$_4$, NH$_3$, CO, N$_2$ and possibly CO$_2$). The remaining constituents consist of silicates (Mg, Si and O-rich material) and iron (as
mixtures of more refractory elements under the form of metal, oxide, sulfide or substituting for Mg in
the silicates). 
In the pressure-temperature ($P$-$T$) domain
characteristic of planet interiors, elements 
go from a molecular or atomic state in the low-density outermost regions to an ionized, metallic one in the dense inner parts, covering the regime of {\it pressure}-dissociation and ionization. Interactions between molecules, atoms, ions and electrons are
dominant and degeneracy effects for the electrons play a crucial role, making the derivation of
an accurate equation of state (EOS) a challenging task. 
Other phenomena such as phase transition or phase separation may take place in the interior of planets, adding complexity to the problem.

The correct description of the structure and cooling of solar or extrasolar planets thus requires the knowledge
of the EOS and the transport properties of various materials under the aforementioned density and temperature conditions.
In the section below, we summarize the recent improvements in this field, both on the experimental and theoretical fronts, and show that tremendous improvements have been accomplished on both sides within the recent years.

\bigskip
\noindent
\textbf{2.1 Hydrogen and helium equation of state}
\medskip

A lot of experimental work has been devoted to the exploration of hydrogen (or its isotope deuterium)  and helium at high densities, in the regime of pressure ionization.
Modern techniques include laser-driven shock-wave experiments \citep{1998Sci...281.1178C, 2001PhRvL..87p5504C, 2000PhRvL..85.3870M}, pulse-power compression experiments \citep{2004PhRvB..69n4209K} and
convergent spherical shock wave experiments \citep{2002JETPL..76..433B, 2003DokPh..48..553B}. They achieve pressures of several Megabars in fluid deuterium at high temperature, exploring
for the first time the regime of pressure-dissociation. For years, however, the difference between the
different experimental results had gave rise to a major controversy. While the laser-driven experiments were predicting significant compression of D along the Hugoniot curve, with a maximum compression factor of $\rho/\rho_0\simeq 6$, where $\rho_0=0.17\,\gcc$ is the initial density of liquid D at 20 K, the pulse experiments were predicting significantly stiffer EOS, with $\rho/\rho_0\simeq 4$. 
The controversy has been resolved recently thanks to a new determination of the quartz EOS up to 16 Mbar  \citep{2009PhRvL.103v5501K}. Indeed, evaluation of the properties of high pressure material is made through comparison with a shock wave standard.
Quartz is the most commonly used reference material in the laser experiments. The new experiments have shown that, due to phase changes at high density, the quartz EOS was significantly stiffer than
previously assumed. The direct consequence is thus a stiffer D and H EOS, now in good agreement with the pulse-power compression data.  
 
 Except for the release of the new quartz EOS (see above) which led to a reanalysis of the existing data, no new result has been obtained for H on the experimental front for more than a decade. Because of the high diffusivity of this material, experimentalists have elected to focus on helium, more easy to lock into pre-compressed cells.
Furthermore, as helium pressure ionization has been predicted to occur directly from the atomic (He) state into the fully ionized (He$^{++}$) one \citep{WinisdoerfferChabrier2005}, this material holds the promise to bring valuable information on the very nature of pressure metallization. 
For this reason, even though He by itself represents only 10\% of giant planet interior compositions, against 90\% for H, all the progress on the high pressure EOS of light elements since the PPV has been obtained for this material, as reviewed below. 

 Recent high-pressure experiments, using statically precompressed samples in dynamical compression experiments, have achieved up to 2 Mbar for various Hugoniot initial conditions, allowing to test the EOS over a relatively broad range of $T$-$P$ conditions \citep{Eggert08}. These experiments seem to show a larger compressibility than for H, 
 possibly due to electronic excitations, in good agreement with the SCVH EOS \citep{SCVH}. These results, however, should be reanalyzed carefully before
 any robust conclusion can be reached. Indeed, these experiments were still calibrated on the old quartz EOS. It is thus expected that using the recent (stiffer) one will lead to a stiffer EOS for He, as for H, in better agreement with ab-initio calculations \citep{Militzer06}. 

Tremendous progress has also been accomplished on the theoretical front.
Constant increase in computer performances now allows
 to perform
\textit{ab initio} simulations over a large enough domain of the $T-P$ diagram to generate appropriate EOS tables.
These approaches include essentially
quantum Molecular Dynamics (QMD) simulations, which
combine molecular dynamics (MD) to calculate the forces on the (classical) ions and Density Functional Theory (DFT) to take into account the quantum nature of the electrons.
We are  now in a position where the semi-analytical 
Saumon-Chabrier-Van Horn (SCVH) EOS for H/He can be supplemented by first-principle EOS \citep{2011PhRvB..83i4101C,2012ApJ...750...52N,2013PhRvB..87a4202M} ({\it Soubiran et al. in prep.}).

A point of importance concerning the H and He EOS is that, while the experimental determination of
hydrogen pressure ionization can be envisaged in a foreseeable future, reaching adequate pressures for helium ionization remains presently out of reach. One way to circumvent this problem is to explore the dynamical (conductivity) and optical (reflectivity) properties of helium at high density. Shock compression conductivity measurements had suggested that He should become a conductor at $\sim$ 1.5 g cm$^{-3}$ \citep{2003JETP...97..259F}. The conductivity was indeed found to rise rapidly slightly above $\sim$ 1 g cm$^{-3}$, with a very weak dependence upon
temperature, a behavior attributed to helium pressure ionization (alternatively the closure of the band gap). Note, however, that the reported measurements are model dependent and that the conductivity determinations imply some underlying EOS model, which has not been probed in the domain of interest.
Besides the conductivity, reflectivity is another diagnostic of the state of helium at high density.
Reflectivity can indeed be related to the optical conductivity through the complex index of refraction, which involves the contribution of both bound and charge carrier electron concentrations.
Optical measurements of reflectivity along Hugoniot curves were recently obtained in high-pressure experiments, combining static and shock-wave high-pressure techniques, reaching a range of final densities $\sim 0.7$-1.5 g cm$^{-3}$ and temperatures $6\,10^4$ K \citep{2010PhRvL.104r4503C}. Within the regime probed by
the experiments, helium reflectivity was found to increase continuously, indicating
increasing ionization. A fit to the data, based on a semiconductor Drude-like model,
was also predicting a mobility
gap closing, thus helium metallization, around $\sim$ 1.9 g cm$^{-3}$ for temperatures below $T\lesssim 3\,10^4$ K.
These results were quite astonishing. It is indeed surprising that helium, with a tightly bound, closed-shell electronic structure and an ionization energy of 24.6 eV, would ionize at such a density, typical of hydrogen ionization.

All these results were questioned by QMD simulations which were predicting exactly the opposite trend, with a weak dependence of conductivity upon density and a strong dependence on temperature \citep{2007PhRvB..76g5112K}. The QMD calculations found that reflectivity remains very small at a density $\sim 1$ g cm$^{-3}$ for $T\lesssim 2$ eV, then rises rapidly at higher temperatures, due to the reduction of the band gap and the
increasing occupation of the conduction band arising from the (roughly exponential) thermal excitation of electrons.
Accordingly, the calculations were predicting a band gap closure at significantly higher densities.  Indeed, the band gap energy found in the QMD simulations
($E_{\rm gap}\sim 15$-20 eV) is found to remain much larger than $kT$ in the region covered by the experiments (around $\sim$ 1 g cm$^{-3}$ and 1-2 eV).

This controversy between experimental results and theoretical predictions was resolved recently by QMD simulations, similar to the ones mentioned above but exploring a larger density range \citep{2012PhRvB..86k5102S}. 
These simulations were able to reproduce the experimental data of \cite{2010PhRvL.104r4503C} with
excellent accuracy, both for the reflectivity and the conductivity, but with a drastically different predicted behavior of the gap energy at high density. Based on these results, these authors revisited the  experimental ones. They used a similar Drude-like model to fit their data, with a similar functional dependence for the gap energy on density and temperature, but with a major difference. While \cite{2010PhRvL.104r4503C} ignored the
temperature dependence of the gap energy, keeping only a density-dependent term, \cite{2012PhRvB..86k5102S} kept both terms. As a result, these authors find a much lower dependence on density
than assumed in the previous experimental data analysis and a strong dependence on temperature.
This leads to a predicted band closure, thus He metallization, in the range $\sim 5$-10 g cm$^{-3}$ around 3 eV. Interestingly, this predicted ionization regime is in good agreement with theoretical predictions of the He phase diagram \citep{WinisdoerfferChabrier2005}. Importantly enough, the regime $\rho \sim 2$-4 g cm$^{-3}$, $T\lesssim 10^4$ K should be accessed in a near future with precompressed targets and drive laser energies $\ge 10$ kJ. Such experiments should thus be able to confirm or reject the theoretical predictions, providing crucial information on the ionization state of He at high densities, characteristic of giant planet interiors.


\bigskip
\noindent
\textbf{2.2 Heavy elements}


\medskip\noindent
{\it Iron} 
\smallskip

Current diamond anvil cell experiments reach several thousands degrees at a maximum pressure of about 2 Mbar for iron \citep{1993Natur.363..534B}, still insufficient to explore the melting curve at the Earth inner core boundary ($\sim 3$ Mbar and $\sim$ 5000 K). On the other hand, dynamic experiments yield too high temperatures to explore the relevant $P$-$T$ domain for the Earth but may be useful to probe e.g. Neptune-like exoplanet interior conditions. So far, one must thus rely on simulations to infer the iron melting curve for Earth, super-Earth and giant planet conditions.
Applying the same type of QMD simulations as mentioned in the previous sections, Morard et al. ({\it Morard, Bouchet, Mazevet, Valencia, Guyot, EPSL submitted}) have determined the melting line of Fe up to $T=$ 14000 K and $P=$15 Mbar. For the Earth, although this melting line lies above the
typical temperature at the core mantle boundary, adding up $\sim 10$-30\% of impurities would lower
the melting temperature enough to cross this boundary. The very nature of iron alloy at the Earth inner core boundary thus remains presently undetermined. In contrast, the melting line, even when considering the possible impact of impurities, lies largely above the $T$-$P$ conditions
characteristic of the core mantle boundary for super-Earth exoplanets ($M\sim 1$-10 M$_\oplus$).
It is thus rather robust to assert that the iron core of these bodies should be in a solid phase, precluding the generation of large magnetic field by convection of melted iron inside these objects.
These calculations should be extended to higher temperatures and pressures in a near future to explore the melting line of iron under Jovian planet conditions.

\medskip\noindent
{\it Water, rocks} 
\smallskip

The most
widely used EOS models for heavy elements are ANEOS \citep{Thompson72} and SESAME \citep{Lyon92}, which describe the thermodynamic properties of water, "rocks" (olivine (fosterite Mg$_2$SiO$_4$) or dunite in ANEOS, a mixture of silicates and other heavy elements called "drysand" in SESAME) and iron. These EOS consist of
interpolations between models calibrated on existing Hugoniot data, with thermal corrections approximated by a Gr\"uneisen parameter ($\gamma=\frac{V}{C_V}(\frac{dP}{dV})_T$), at low to moderately high
($\simle 0.5$ Mbar) pressure, and Thomas-Fermi or more sophisticated first-principle calculations at very
high density ($P\simgr 100$ Mbar), where ionized species dominate. Interpolation between these limits, however, provides no insight about the correct structural and electronic properties of
the element as a function of pressure, and thus no information about its compressibility, ionization stage (thus conductibility), or even its
phase change, solid or liquid. All these properties can have a large impact on
the internal structure and the evolution of the planets (see \S 5.1). 

Recent high-pressure shock compression experiments of unprecedented accuracy for water, however, show significant  departures from both the SESAME and ANEOS EOS in the $T$-$P$ domain characteristic of planetary interiors, revealing a much lower compressibility \citep{Knudson12}. 
This is consequential for giant planet interiors, in particular Uranus or Neptune like planets.
Indeed, as the calculated amount of H and He in the planet decreases with the stiffness of the water EOS, this new EOS suggests the presence of some H/He fraction in the deep interior of Neptune and Uranus, excluding an inner envelope composed entirely of "water like" material \citep{Fortney10}. As H would be metallic, this bears important consequences for the generation of the magnetic field, as addressed below.

The new data, on the other hand, are in excellent agreement with QMD simulations \citep{French09}. This again reinforces confidence in EOS calculated with first-principle methods and we are now in a state to be able to use these latter to compute reliable EOS for water or other material in the
density regime relevant for planetary interiors \citep[][{\it ; Licari et al., in prep.}]{French09, 2012ApJS..202....5F}.

Several major predictions emerge from all the aforementioned first-principle calculations devoted to the EOS of water at high pressure and density. 
First of all, all these simulations predict a stable superionic phase for water at high density,  characterized by mobile protons in an icy structure \citep{2011Icar..211..798R, Wilson12, 2013PhRvL.110o1102W}.
While water is always found to be in a liquid, plasma state under the conditions of Jupiter's core ($\sim$ 20000 K, 50 Mbar), 
it should be in the superionic state in a significant fraction of Neptune and Uranus inner envelope.
For Saturn's, its state is more uncertain, the uncertainties on the exact $T$-$P$
core-envelope boundary conditions for this planet encompassing the predicted phase transition line.
Besides having a significant impact on the determination of the very nature of the core inside solar or extrasolar giant planets, including the so-called "ocean planets" predicted to have large water envelopes, these calculations bear major consequences on the generation of the magnetic fields in Uranus and Neptune,
with unusual non-dipolar components, in contrast to that of the Earth.

\bigskip
\noindent
\textbf{2.3 Phase separation}
\medskip

The existence of a phase separation between hydrogen and helium under conditions characteristic of Jupiter and Saturn interiors, suggested several decades ago \citep{1973ApJ...185L..95S, 1973ApJ...181L..83S}, remains an open problem. Such a phase separation is suggested by the measurement of atmospheric He abundance in Jupiter (see \S 4.1).
Under the action of the planet's gravity field, a density discontinuity yields an extra source of gravitational energy as the dense phase droplets  (He-rich
 ones in the present context) sink towards the planet's center. Conversion of this gravitational energy into heat delays the cooling of the planet,
 which implies a larger age to reach a given luminosity compared with a planet with a homogeneous interior.
 In Saturn's case, such an additional source of energy has traditionally been suggested to explain the planet's bright luminosity at the correct age, i.e. the age of the Solar System, $\sim 4.5\times 10^9$ yr \citep{1977ApJS...35..221S}, although an alternative explanation has recently been proposed by \cite{Leconte13} (see \S 4). 
 
 Recently, two groups have tried to determine the shape of the H/He phase diagram under Jupiter and Saturn interior conditions, based again on QMD simulations for the H/He mixture under appropriate conditions \citep{2009PNAS..106.1324M, 2011PhRvB..84w5109L}.
 Although basically using the same techniques and reaching globally comparable results, the H/He phase diagram and critical line for H/He for the appropriate He concentration obtained in these two studies
 differ enough to have significantly different consequences on Jupiter and Saturn's cooling histories.
 Indeed, it must be kept in mind that, even if H/He phase separation does occur for instance in Saturn's interior, not only it must encompass a large enough fraction of the planet interior for the induced gravitational energy release to be significant, but it must occur early enough in the planet's cooling history for the time delay to be consequential today. This implies rather strict conditions on the shape of the phase diagram \citep[see e.g][]{2004ApJ...608.1039F}. The critical curve obtained by \cite{2009PNAS..106.1324M}, for instance, does not yield enough energy release to explain Saturn's extra luminosity and excludes H/He in Jupiter's present interior. In contrast, the critical line obtained by \cite{2011PhRvB..84w5109L} predicts a phase separation to take place inside both Jupiter and Saturn's present interiors and roughly fulfills the required conditions to explain Saturn's extra luminosity.
This does not necessarily imply that one of the two diagrams, if any, is correct and the other one is not. Indeed, as shown by \cite{Leconte13}, if layered convection occurs within Saturn's interior, it can explain all or part of the excess luminosity (see \S 4.4), making the contribution of a possible
 phase separation less stringent. Clearly, the issue of the H/He phase diagram under giant planet interior conditions and its exact impact on the planet cooling remains presently an unsettled issue.
 
 To complement this study, recent similar QMD calculations have focused on the possible optical signature of an equimolar or near equimolar H/He mixture undergoing demixtion \citep{2011PhRvB..84p5110H, 2013PhRvB..87p5114S}. These studies have calculated the expected change of reflectivity upon demixing. It is predicted that reflectivity exhibits a distinctive signature between the homogeneous and demixed phases potentially observable with current laser driven experiments \citep{2013PhRvB..87p5114S}.
  
Of notable interest concerning immiscibility effects in jovian planet interiors are also the recent results of \cite{Wilson12, Wilson12b}. Performing QMD simulations, these authors have shown that for pressures and temperatures characteristic of the core envelope boundary in Jupiter and Saturn, water and MgO (representative of rocky material) are both soluble in metallic hydrogen. This implies that the core material in these planets should dissolve into the fluid envelope, suggesting the possibility for significant core erosion and upward redistribution of heavy material in Jovian solar and extrasolar planets. A non uniform heavy element distribution will limit the rate at which the heat flux can be transported outwards, with substantial implications for the thermal evolution and radius contraction of giant planets (see \S 4.4 and \S 5.2).

\bigskip
 
\centerline{\textbf{ 3. STRUCTURE OF TERRESTRIAL PLANETS}}
\bigskip

	A terrestrial exoplanet is defined as an exoplanet having characteristics similar to the Earth, Mars and Venus. 
	The Earth is the only planet where the existence of life has been demonstrated. The search for terrestrial exoplanets is driven by the question of life on alien worlds and the belief that it will be on a planet that resembles the Earth. The information available for characterizing exoplanets is limited to the distance to its star, the radius, the mass, and very rarely some information on the atmospheric composition \cp[e.g][]{Swain08}. This section is devoted to the description of the interior structure and dynamics of terrestrial planets, starting with the Earth, which is the best known terrestrial planet. One unique characteristic of the Earth is that its surface is divided into several rigid plates that move relative to one another. The motion of the plates is driven by thermal convection in the solid mantle \cp{Schubert01}. This convection regime is known as the mobile lid regime. Most of the Earth's volcanism happens at the plate boundaries. Plate tectonics provide a recycle mechanism that may be important for sustaining life on a planet although this is not demonstrated. It is also a very efficient way to cool down a planet compared to the so-called 'stagnant-lid' regime (no plate tectonics) that characterizes both Venus and Mars. After describing the interior structure of a terrestrial planet (\S 3.1), this chapter describes the interior dynamics and addresses the question of the link between convection and plate tectonics (\S 3.2). Some small exoplanets with a radius less than two Earth-radius may be terrestrial exoplanets. The last section briefly describes some of them and shows that none of them can have liquid water on its surface.
	
\bigskip
\noindent
\textbf{3.1 The interior structure of terrestrial planets}

\medskip\noindent
{\it The elements that form a terrestrial planet} 
\smallskip

\begin{figure}[h]
\vspace{-7.2cm}
\hspace{-6.5cm}
  \includegraphics[scale=0.75]{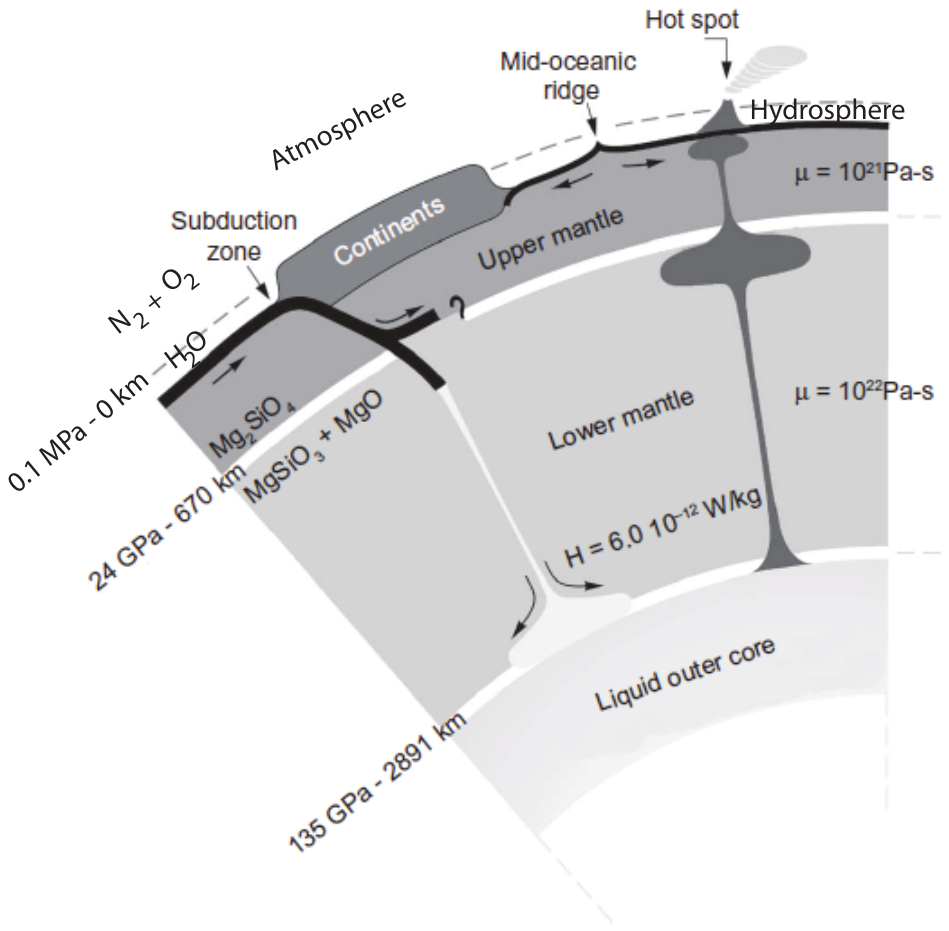}
  \vspace{-8.cm}
  \caption{\small Structure of the Earth. 
 Seismic data, laboratory experiments, and numerical simulations are used in order to simulate the interior structure and dynamics of Earth. Pressure and depth of major interfaces are indicated on the left. The elements and molecules which compose each layer are described in the text. Adapted from \ct{Sotin11}.
}  \label{fig1_CS}
 \end{figure}

Earth is the terrestrial planet for which we have the most information about its elementary composition. The different layers that form the Earth are from the outside to the inside the atmosphere, the hydrosphere, the crust, the mantle, the liquid core and the solid core (Fig. \ref{fig1_CS}). The formation of the different layers is the result of differentiation processes that started during the accretion 4.5 Gyr ago and are still operating at the present time.

\begin{figure}[h!]
\vspace{-3.6cm}
\hspace{-0.5cm}
\includegraphics[scale=0.4]{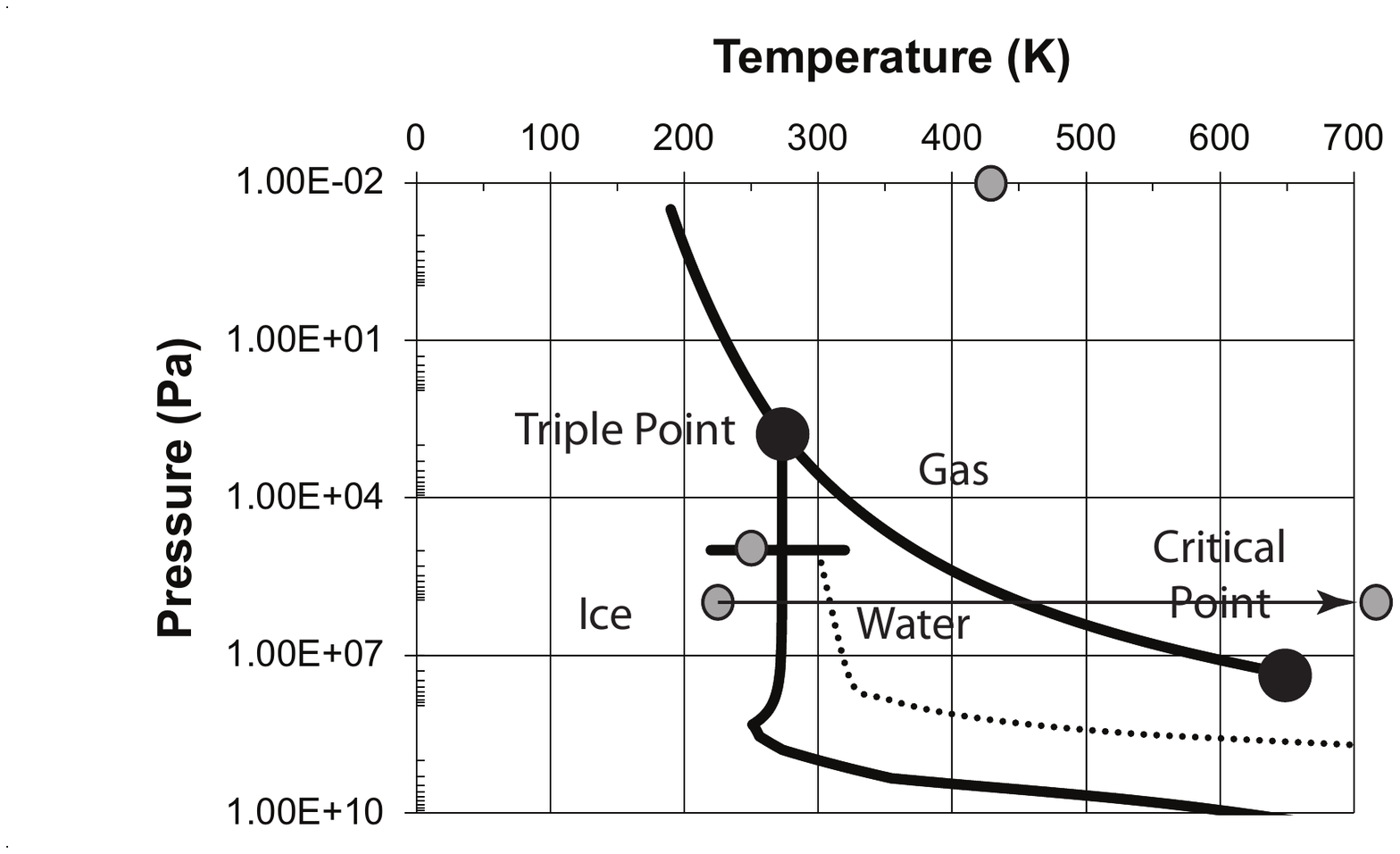}  
\vspace{-4.cm}
  \caption{\small Phase diagram of H$_2$O, indicating the small temperature range for the stability of water at the surface of an exoplanet. The horizontal bar represents the ($P,T$) range at the Earth's surface. The grey circles give the effective temperatures for the Earth, Venus, and Mercury and are placed at the surface pressure of those planets. For Venus, the arrow links the effective temperature to the surface temperature, stressing the role of the atmosphere. The temperature profile inside the Earth (dotted line) lies within the stability domain of liquid water. If the surface pressure is too small, above the triple point (Mars, Mercury), or if the temperature is too hot, above the critical point (Venus), then the liquid water is not stable at the surface.
}  \label{fig2_CS}
 \end{figure}

The atmosphere of the Earth is mainly composed of N and O. Its mass is less than 10$^{-6}$ $\mearth$, which is quite negligible. The hydrosphere (liquid H$_2$O) is important for life to form and to develop. Its mass is about 2 10$^{-4}$ $\mearth$. Having a stable liquid layer at the surface of a planet is a characteristic shared only with Titan, Saturn's largest moon, where liquid hydrocarbons form seas and lakes \cp{Stofan07}. The ($P, T$) stability domain of the liquid layer is very limited in temperature (Fig. \ref{fig2_CS}) and the conditions for liquid water to be present at the surface of a planet impose strong constraints on the atmospheric processes. Two different kinds of crust are present at the surface of the Earth: the oceanic crust that is dense, thin (6 km), and continuously formed at mid-ocean ridges and recycled into the mantle at subduction zones; and the continental crust which is old (on average), thick, and light. The oceanic crust is formed by melting processes that start deep in the mantle beneath mid-ocean ridges. The crust represents about 0.4\% of the Earth's mass and will be neglected for determining the total mass of a planet. Thus, the three layers that are most important for life to start and evolve (atmosphere, hydrosphere, and crust) represent a very small fraction of Earth's mass and cannot be detected just by having the mass and radius of a planet.

The two most massive layers are the mantle and the core. They differentiated from one another because an iron alloy is denser and melts at lower temperature than the silicates. This differentiation processes may have occurred into the planetesimals by segregation and into protoplanets by Rayleigh-Taylor instabilities \cp{Chambers05}. When the protoplanets collided to form the terrestrial planets, their iron cores would have merged. Due to Earth's seismically active interior, the location of the different interfaces, and the pressure and density profiles are very well known \cp{Dziewonski81}. The mantle represents approximately two-third of the mass and is separated into an upper mantle and a lower mantle. The difference is due to the mineralogical transformation of olivine (Mg,Fe)$_2$SiO$_4$ at low pressure to perovskite (Mg,Fe)SiO$_3$ and magnesowustite (Mg,Fe)O at higher pressure. The rock that composes the upper mantle is known as peridotite and is made of olivine, orthopyroxene (Mg,Fe)$_2$Si$_2$O$_6$, clinopyroxene Ca(Mg,Fe)Si$_2$O$_6$ and garnet which is an aluminium-rich silicate. The core is composed of a solid inner core of pure iron nestled inside a liquid outer core that contains a light element, presumably sulfur, in addition to iron.
 

The elementary composition of the Earth can be restricted to 8 elements that represent more than 99.9\% of the total mass \cp{Javoy95}.
Four of these 8 elements (O, Fe, Mg, Si) account for more than 95\% ot the total mass. The four other elements (S, Ca, Al, Ni) add complexities to the model. As described in \ct{Sotin07}, Ni is present with iron in the core, sulfur can be present in the liquid outer core and form FeS, calcium behaves like magnesium and forms clinopyroxene in the upper mantle and calcium-perovskite in the lower mantle, and aluminium substitutes to Si and Mg in the silicates. Although CI chondrites have a solar composition, it has been suggested that the composition of Earth may be similar to that of EH enstatite chondrites \cp{Javoy95, Mattern05}. In this case, rocks are enriched in silicon and the ratio Mg/Si and Fe/Si are lowered to 0.734 and 0.878, respectively (Table \ref{table2_CS}). However, such variations have minor influences on the values of mass and radius and it seems appropriate to use the stellar composition as a good first-order approximation of the composition of the planet {\cp{Sotin07, Grasset09, Sotin11}. 

The elementary composition of some of the stars hosting exoplanet candidates is known \cp{Beirao05, Gilli06}. \ct{Grasset09} demonstrated that their corrected values of Mg/Si and Fe/Si (adding Ca, Al, Ni to their closest major element) vary between 1 and 1.6, and between 0.9 and 1.2, respectively. The Sun is on the lower end of these ratios, suggesting that it is enriched in silicon compared to these stars. However, such variability should not significantly affect the radius of an exoplanet for a given mass \cp{Grasset09}. 

\begin{table}
\caption{Abundances of magnesium and iron relative to silicon for the solar model and the Enstatite model of \ct{Javoy95}.}
\label{table2_CS}
\begin{tabular}{lllll}
 \hline
 &  Solar$^1$ & EH$^1$ & Solar$^2$ & EH$^2$ \\
Fe/Si & 0.977 &  0.878 & 0.986 & 0.909 \\
Mg/Si & 1.072 & 0.734  & 1.131 & 0.803\\
  \hline
   \end{tabular}

    $^1$ 4 elements (O, Fe, Mg, Si) \\
 $^2$ 8 elements (O, Fe, Si, Mg, Ni, Ca, Al, S) \\
  \end{table}

For terrestrial planets that are larger than Earth, higher pressures will be reached in the interior and additional transformation to more condense phases should occur. For example, a post-perovskite phase has been predicted from ab initio calculations \cp[][and references therein]{Stamenkovic11}. However, very little is known on this phase and its thermal and transport properties are still debated \cp{Tackley13}. The elements that compose these very high-pressure minerals shall remain the same.

\medskip\noindent
{\it The equation of state for the different layers} 
\smallskip

The most accessible information about exoplanets is mass and radius. Following the pioneering work of \ct{Zapolsky69} several studies have looked at the relationships between radius and mass \cp{Valencia06, Sotin07}. The mass of a planet is integrated along the radius assuming that a planet can be described as a one-dimensional sphere, i.e. density depends only on radius and does not vary significantly with longitude or latitude. The density at a given radius depends on pressure and temperature and is computed using EOSs which have been obtained for most of the pressure range relevant for Earth due to recent progress in high-pressure experiments \cp[see e.g][and \S 2]{Angel09}. The pressure is determined assuming hydrostatic equilibrium. The temperature is calculated assuming thermal convection in the mantle and core. The temperature gradient is adiabatic, except in the thermal boundary layers \cp[e.g][]{Sotin11}. 

Two different approaches are commonly used in Earth sciences for describing the pressure and temperature dependences of materials \cp[e.g][]{Jackson98}. 
One method introduces the effect of temperature in the parameters that describe the mineral's isothermal EOS and is achieved using the 3rd order Birch-Murnaghan EOS with the thermal effect incorporated using the mineral's thermal expansion coefficient. The second approach dissociates static pressure and thermal pressure by implementing the Mie-Gr\"uneisen-Debye (MGD) formulation. The 3rd order Birch-Murnaghan (BM) EOS is usually chosen for the upper mantle where the pressure range is limited to less than 25 GPa by the dissociation of ringwoodite into perovskite and periclase. The Mie-Gr\"uneisen-Debye (MGD) formulation is preferred for the lower mantle and core. Other EOSs can be used such as the Vinet EOS which provides the same result as BM and MGD at low pressure, because the parameters entering into these equations are well-constrained from laboratory experiments. Within the temperature range of the Earth's lower mantle, the thermal pressure term in the Mie-Gr\"uneisen-Debye formulation provides estimates close to EOS derived from ab initio calculations and shock experiments \cp{ANEOS}.
The composition, i.e. the relative amount of silicon, iron, and magnesium, does not play a major role within the variability of chemical models for the Earth. Whether one takes the EH model or the solar/chondritic model (Table \ref{table2_CS}), the value of the radius remains within the error bars of radius determination for a given mass. The stellar composition variability is of the same order as the variability in composition between the EH model and the chondritic model \cp{Grasset09}. It seems therefore reasonable to use the stellar composition (Fe/Si and Mg/Si) as a reasonable guess of the planet elementary composition. Then the relative amount of each mineral can be calculated as described in \ct{Sotin07}.

Super-Earths are more massive than Earth and the validity of these equations at much higher pressures and temperatures is questionable \cp[see \S2;][]{Grasset09, Valencia09}. The Vinet and MGD formulation appear to be valid up to 200 GPa \cp[e.g][]{Seager07}. Above ~200 GPa, electronic pressure becomes an important component which cannot be neglected. At very high pressure ($P$ $>$ 10 TPa), first principles EOS such as the Thomas-Fermi-Dirac (TFD) formulation can be used \cp[see \S 2;][]{Fortney07a}. The pressure at the core-mantle boundary  of terrestrial planets 5 and 10 times more massive than Earth is equal to 500 GPa and 1 TPa, respectively \cp{Sotin07}. In this intermediate pressure range, one possibility is to use the ANEOS 
EOS mentioned in \S 2.2.
The study by \ct{Grasset09} compares the density-pressure curves of iron and forsterite using the MGD, TFD and ANEOS formulations. The TFD formulation predicts values of densities much too small at low pressure. On the other hand, the ANEOS seems to fit the MGD at low pressure and the TFD at very high pressure. Therefore, the ANEOS appears to be a good choice in the intermediate pressure range from 0.2 to 10 TPa.
If the density was constant with radius, the radius of a planet would vary as M$^{1/3}$.  Since the density increases with increasing pressure and the temperature effect is negligible \cp{Grasset09}, the exponent is actually lower than 1/3 and equal to 0.274 for planets between 1 and 10 Earth-mass. For smaller planets between 0.01 and 1 Earth-mass the coefficient is equal to 0.306, which is closer to 1/3, as expected (Fig. \ref{fig3_CS}). 

\begin{figure}[h]
\vspace{-3cm}
\hspace{-1.4cm}
  \includegraphics[scale=0.45]{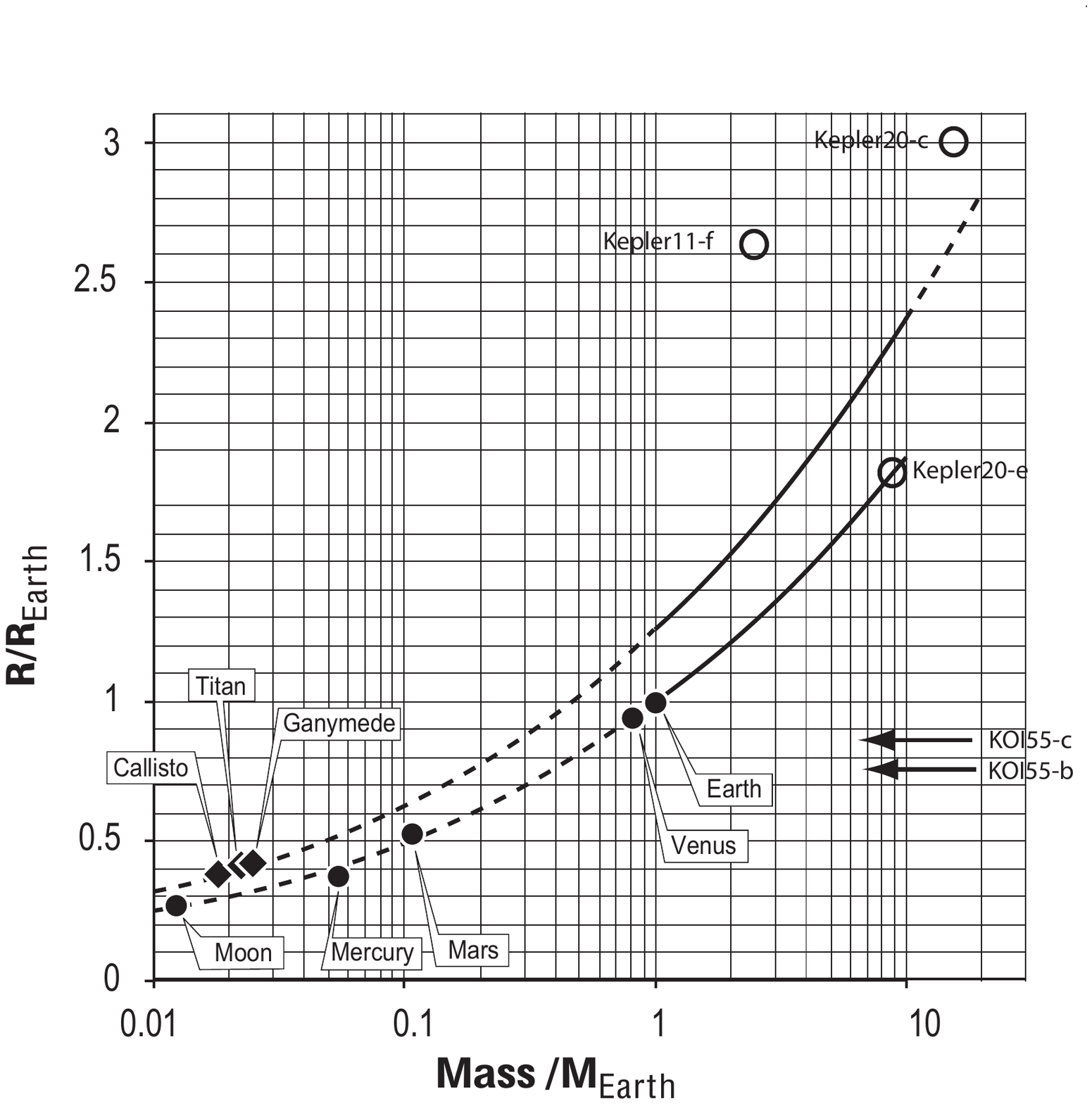}
  \vspace{-3.5cm}
  \caption{\small The Mass-Radius curve for terrestrial planets very well predicts the values observed for the Solar System terrestrial planets. The second curve corresponds to the ($R,M$) curve for planets having 50\% mass of H$_2$O. The radius is about 25\% larger than that of a terrestrial planet. The large icy moons of Jupiter and Saturn fit very well on this curve. The open circles give the values for some small exoplanets (see \S 3.3).
}  \label{fig3_CS}
 \end{figure}
 
\medskip\noindent
{\it Observations} 
\smallskip

Two different observations can be used to test the validity of such models. First, one can compare the mass and radius of the terrestrial planets in our Solar System. Second the density profile of the Earth can be compared with the calculated one.
	The mass and the radius of Earth, Mars, and Venus are very well known. The curve describing how a radius depends on the mass of a terrestrial planet is drawn on Figure \ref{fig3_CS}. The fit is very good with the predicted radius of Mars and Mercury being slightly smaller and larger, respectively, than the measured value. For Mercury, it is well known that the ratio Fe/Si is much larger than for the other terrestrial planets. Having more iron reduces the size of the planet. It is therefore not surprising that the radius calculated for a solar-type composition is too large. However, the difference is quite small compared to the uncertainties attached to the determination of the radius and mass of exoplanets. For Mars, the difference is also very small. Mars may have less iron. Future measurements obtained by the INSIGHT mission that will carry a seismometer on the surface of Mars will help resolve that issue. For Venus and the Earth, the calculated value is equal to the observed one at less than 1\%. For example, for the Earth the calculated value is 43 km larger than the observed value \cp{Sotin07}. For Venus whose mass is 0.81 Earth-mass, the calculated radius is only 5 km larger. The simple model using solar composition for eight elements provide a very good approximation of the observed (mass, radius) of the terrestrial planets in our Solar System.
	
	The earthquakes generated by the motion of the plates have allowed a precise description of the density structure of the Earth \cp{Dziewonski81}. This density profile can be compared with the calculated one (Fig. \ref{fig4_CS}). The major difference is in the density of the inner core because the simple model does not model the crystallization state of the core. On Earth, the inner core is solid and made of iron and nickel without any light elements that remain in the liquid phase. Including the growth of an inner core in the model adds complexities that are not required because it does not change significantly the radius for a given mass. The density profile is different in the inner core but information on the presence of an inner core in exoplanets is not yet available. The Earth's inner core only represents one tenth of the Earth's core in mass, which can be translated into about 4\% of the total mass of the Earth. Besides the inner core, the agreement between the calculated density and the observed density is very good and one can barely see the difference. Another place where the two curves differ is at the transition between the lower and the upper mantle. In the model, the difference between the two mantles is an abrupt change due to the transformation of olivine into perovskite and magnesowustite. However, we know from seismic observations and laboratory experiments that there is a transition zone in which the olivine transforms into spinel phases. This complexity is not included in the model since it does not affect the value of the radius for a given mass.

 \begin{figure}[h]
 \vspace{-8cm}
 \hspace{-3.5cm}
  \includegraphics[scale=0.7]{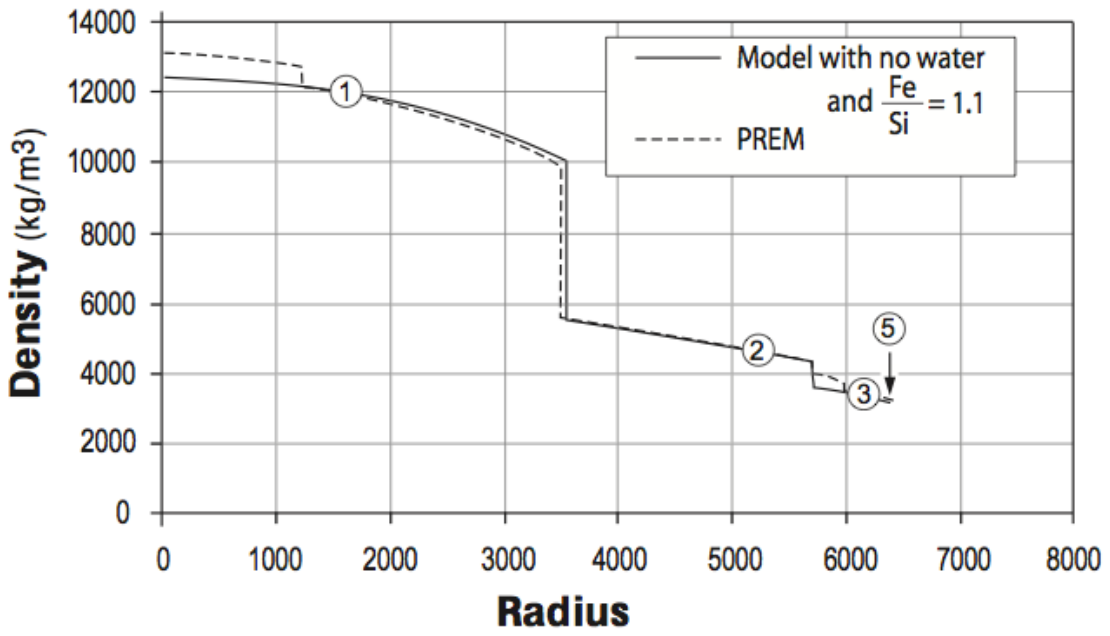}
  \vspace{-9cm}
  \caption{\small The density profile simulated by the simple Earth model compares very well with the Preliminary Reference Earth Model (PREM) determined by inversion of seismic waves propagating through the Earth. The simple model of the Earth has only three layers: the core (1), the lower mantle (2), and the upper mantle (3). The layer (5) is the hydrosphere which represents less than 0.1\% of the total mass. The model and the observation are somewhat different for the inner core because the inner core is made of pure iron whereas the model includes the sulfur in both the solid inner iron core and the liquid outer iron-rich layer.
}  \label{fig4_CS}
 \end{figure}

\bigskip
\noindent
\textbf{3.2 The internal dynamics}
\medskip

One unique feature of the Earth is the presence of plate tectonics. Although Venus has global properties very similar to Earth, its surface does not present any sign of such tectonics. The reason(s) why two very similar planets have evolved along very different convective regimes is not known \cp{Moresi98, Stevenson03}. The next section provides an overview of mantle convection followed by a summary of some recent numerical simulations on the relationships between convection and plate tectonics. 

\medskip\noindent
{\it Subsolidus convection in the silicate mantle} 
\smallskip

The convection pattern in the Earth's mantle is characterized by hot plumes rising from the core-mantle boundary towards the surface. They stop at the cold boundary layer some tens of kilometers below the surface because the upper part of the mantle has a much too high viscosity (see below). They eventually form hot spots. Cold sheet of oceanic lithosphere dive into the mantle at subduction zones (Fig. \ref{fig1_CS}). Thermal convection is much more efficient than conduction to remove heat from the interior of a terrestrial planet. Heat sources include radiogenic internal heating due to the decay of the long-lived radioactive elements $^{40}$K, $^{232}$U, $^{235}$U and $^{242}$Th, and the initial heat stored into the planet during the accretion and the differentiation. The convective processes in the mantle control the thermal evolution of the planet (e.g. Schubert et al., 2001). In a fluid which is heated from within, cooled from the top (cold surface temperature) and heated from below (hot core), cold plumes form at the upper cold thermal boundary layer and hot plumes form at the hot thermal boundary layer which corresponds to the core-mantle boundary \cp[e.g.][]{Sun07}. The efficiency of heat transfer is mainly controlled by the mantle viscosity which depends on a number of parameters including, but not limited to, the mineral composition of the mantle, the temperature, the pressure and the grain size. Laboratory experiments \cp{Davaille93} and numerical studies \cp[e.g][]{Moresi98, Grasset98} have stressed the major role of temperature-dependent viscosity. The vigor of convection is measured by the Rayleigh number that represents the ratio between the buoyancy forces (density variations induced by temperature) and the viscous force. A small viscosity leads to small viscous forces and large values of the Rayleigh number. Scaling laws have been derived to express the heat flux as a function of the Rayleigh number. A high value of the Rayleigh number leads to a high value of the heat flux. These laws have been employed to predict the thermal evolution of exoplanets \cp[e.g][]{Valencia07}. Such scaling laws are valid in the stagnant lid regime.

One characteristic of the stagnant lid regime is the presence of a thick conductive layer above the convective mantle. The presence of this layer is due to the very strong temperature-dependence of viscosity \cp{Davaille93}. At the surface where the temperature is cold, the material has a viscosity that is several orders of magnitude larger than the viscosity of the convective mantle. This layer is known as the lithosphere. The cold thermal boundary layer where cold plumes can form is located under this stagnant lid. Upwelling plumes can deform that layer and provide a topographic signal as it is the case for Venus \cp{Stofan95}. The adiabatic decompression of the material contained in the uprising plume may lead to partial melt at shallow depths. The partial melt eventually migrates to the surface and causes volcanism that releases gases that were present in the mantle into the atmosphere.

The formation of hot plumes requires the presence of a thermal boundary layer at depth, most likely at the core-mantle boundary. It means that the core is hotter than the mantle, which is possible if the mantle can cool down as quickly as the core. Previous studies based on scaling laws describing the thermal evolution of the core and the mantle \cp[e.g][]{Stevenson83} show that the temperature difference between the core and the mantle decreases quickly. Such a decrease may stop the formation of hot plumes and the convection pattern would be characterized by cold plumes sinking into the mantle and global upwelling as it is the case for a fluid heated from within and/or cooled from above  \cp[e.g][]{Parmentier94}. Another consequence would be the shutdown of the magnetic dynamo \cp{Stevenson83} which has implications on the escape rate of the atmosphere since the presence of a magnetosphere is thought to protect the atmosphere. Mars and Venus do not have a magnetic dynamo at present time. Mars had one during its early history and the characteristics of this magnetic dynamo are recorded in old crustal rocks that contain a remanent magnetization \cp{Acuna99}. It is also thought that the presence of a magnetic dynamo on Earth is related to the more efficient cooling rate of the mantle induced by the plate tectonics regime. Indeed, plate tectonics is more efficient than stagnant-lid convection to remove the heat and to cool down the mantle such that the temperature difference between the mantle and the core enables convective motions in the core. 

Mantle convection is an important process which influences the evolution of the surface and the atmosphere. Critical to our understanding of the evolution of a terrestrial is the relationships between convection and plate tectonics.   

\medskip\noindent
{\it Relationships between convection and plate tectonics } 
\smallskip

Plate tectonics imply that the lithosphere can be broken. As described in \ct{Moresi98}, the lithosphere breaks when the stresses induced by convection become larger than the yield stress. This approach was used by \ct{ONeill07} to simulate the transition from the stagnant-lid regime to the mobile-lid regime, using the Earth's case to scale the transition to planets of different sizes. On the other hand, \ct{Valencia07} used scaling laws that provide relationships between the convective stresses and the Rayleigh number \cp{Schubert01}. The two studies reach two opposite conclusions. On the one hand, \ct{ONeill07} concluded that  increasing planetary radius acts to decrease the ratio of driving to resisting stresses, and thus super-sized Earths are likely to be in an episodic or stagnant lid regime. The episodic regime occurs when the planet experiences episodes of mobile-lid regime. On the other hand, \ct{Valencia07} concluded that as planetary mass increases, the shear stress available to overcome resistance to plate motion increases while the plate thickness decreases, thereby enhancing plate weakness. \ct{Sotin11} pointed out that the two studies can be reconciled if the proper definition of the boundary layer is taken into account. The size of a planet may not be the key parameter for the transition from stagnant lid regime to mobile lid (plate tectonics) regime. More important is the value of the yield strength. 

The yield strength of the lithosphere depends on its history. If the lithosphere has been weakened by impacts, then the yield strength may be much smaller. Also, if the surface temperature is larger, the deformation of the lithosphere may prevent global faulting. \ct{Lenardic12} have developed a model of coupled mantle convection and planetary tectonics to demonstrate that history dependence can outweigh the effects of a planet's energy content and material parameters in determining its tectonic state. This conclusion was already mentioned by \ct{Stevenson03} to explain the different paths followed by the Earth and Venus. The tectonic mode of the system is then determined by its specific geologic and climatic history. The study by \ct{Lenardic12} concludes that models of tectonics and mantle convection will not be able to uniquely determine the tectonic mode of a terrestrial planet without the addition of historical data. It points towards a better understanding of how climate (surface temperature) can influence the tectonic mode. The coupling between a radiative transfer model of the atmosphere and an internal dynamics model is therefore required. The atmospheric model should include the effects of greenhouse gases such as H$_2$O and CH$_4$ and would provide the boundary conditions (pressure, temperature) to the internal dynamics model. The internal dynamics model would determine how much gases can be extracted from the mantle into the atmosphere and would provide the surface heat flux. Such models are not yet available.

\medskip\noindent
{\it Melting in the mantle} 
\smallskip

Melting is a critical process that is responsible for both the formation of the iron-rich core and volcanism that transfers gases dissolved in the mantle to the atmosphere. 
Inspection of the melting curves of iron, iron alloys (Fe-FeS system), and silicates \cp[see Fig. 3 in][]{Sotin07} shows that the melting temperature of a Fe-FeS system is lower than the melting temperature of the silicate mantle.
It implies that an iron rich liquid would form and migrate towards the center of the planet because its density is very large \cp[e.g][]{Ricard09}.

	The solidus of the silicates is above the horizontally averaged temperature profile in the Earth's mantle. The difference between the temperature profile and the solidus decreases with decreasing pressure. Melting of silicates occurs close to the surface (around 100 km) in the hot plumes. This partial melt is less dense and can migrate into magma chambers. Then, the melt contained in the magma chambers eventually reaches the surface (volcanism). Gases play an important role because the solubility of gases decreases with decreasing pressure. Therefore, as the magma migrates towards the surface, the gases exsolve and occupy a larger volume, creating more buoyancy. This runaway effect causes the magmatic eruptions and gases are expelled into the atmosphere. The role of surface pressure is important since it may limit the intensity of the eruptions and the amount of gases released in the atmosphere.


	H$_2$O is an important gas. First it is a greenhouse gas that will increase the surface temperature. Second, liquid water is thought to be a key ingredient for life. Its presence on the surface is limited to a narrow range of temperature (Fig. \ref{fig2_CS}). The evolution of the surface temperature with time as a planet differentiates by volcanism is still poorly understood. The role of plate tectonics is important because it recycles some of the water into the mantle. The water also reacts with the silicates to form hydrated silicates that have properties, including yield strength, very different from those of dry silicates. Modeling the H$_2$O cycle of an exoplanet is required to understand whether it can be similar to Earth and harbor life.

\bigskip
\noindent
\textbf{3.3 Has a terrestrial exoplanet been found?}
\medskip

If the definition of a terrestrial planet is limited to its size and mass, the answer to this question is probably yes. For example the characteristics of the planet Kepler-20b lie upon the silicate (Mass, Radius) curve \cp{Gautier12, Fressin12}. However, if additional conditions, such as surface temperature and atmospheric composition, are required for defining the terrestrial nature of an exoplanet, then the answer is negative. For Kepler-20e (see below), the equilibrium temperature is on the order of 1200 K suggesting the presence of molten silicates at shallow depth if not at the surface.

The Kepler mission has provided about a dozen planets with a radius lower than twice the Earth radius. The two smallest ones are in the Kepler-42 system. Also known as the KOI (Kepler Object of Interest) 961 system, the Kepler-42 system is composed of three planets orbiting a M dwarf with orbital periods of less than two days \cp{Muirhead12}. There is no information about their mass, which makes impossible the determination of their density. Although they orbit an M dwarf, their close distance to the star makes the temperature high with values ranging from 800 K to 500 K.
Precise photometric time series obtained by the Kepler spacecraft during a little less than two years have revealed five periodic transit-like signals in the G8 star Kepler 20 \cp{Gautier12, Fressin12}. These observations provide the radius of the five planets which are, in increasing distance from the star, named as 20b, 20e, 20c, 20f, 20d \cp{Gautier12}. Radial velocities measurements provide the mass of Kepler-20b (8.7 $\mearth$) and Kepler-20c (16.1 $\mearth$). The mass and radius of Kepler-20b are consistent with a terrestrial composition \cp{Gautier12}. The mass and radius of Kepler-20c are more consistent with a sub-Neptune composition. A maximum value of the mass has been inferred for Kepler-20d which also makes it consistent with a Neptune-like composition. The two planets Kepler-20e and Kepler-20f are not massive enough to provide a measurable radial velocity on the star. Without the radial velocity measurement, the determination of the mass relies on theoretical considerations \cp{Fressin12} leading to upper and lower bounds:  0.39 $< M_{\rm Kepler-20e}$/$\mearth <$ 1.67 and 0.66 $< M_{\rm Kepler-20f}$/$\mearth  <$ 3.04. These two planets are located further away from the star than Kepler-20b. Therefore, one might expect them to contain more volatiles. Since Kepler-20b lies upon the terrestrial Mass-Radius curve, the planets Kepler-20e and Kepler-20f are likely above although that statement needs radial velocity measurements to be validated. Finally, the equilibrium temperature is quite large and equal to 1136 K and 771 K for Kepler-20e and Kepler-20f, respectively. 

The Kepler-11 system is composed of six planets for which the masses of the five closer to the star have been estimated by their transit time variations \cp{Lissauer11}. Although transit time variations can only provide upper limits for the mass, the values suggest that the densest planet is the closest to the star. However, even with the maximum mass, these planets are less dense than silicate planets.
Lastly, two of the smallest known planets are orbiting the post-red-giant, hot B subdwarf star KIC 05807616 at distances of 0.0060 and 0.0076 AU, with orbital periods of 5.7625 and 8.2293 hours, respectively \cp{Charpinet11}. The radius of these two planets KOI-55b and 55c is equal to 0.759 $\rearth$ and 0.867 $\rearth$, respectively (see Fig. \ref{fig3_CS}). These planets are smaller than Earth. However, there is no constraint on their mass and it is therefore impossible to conclude that they are terrestrial planets. Finally, their equilibrium temperature is larger than 7,000 K. Such high values of the surface temperature do not fit the canonical model of a terrestrial planet.

\bigskip
\noindent
\textbf{3.4 Perspectives}
\medskip

The curve (mass, radius) of terrestrial planets is well determined and has been tested against the Earth's characteristics. Varying the elementary composition does not provide significant variations to the terrestrial curve. The Kepler mission has detected more than a dozen planets with a radius smaller than twice the Earth-radius. However, only a few of them fall upon the terrestrial curve. For some of them, precise determination of the mass is still lacking. Plate tectonics may play a major role in the development of life. Terrestrial planets that resemble the Earth have to be in that regime.
 
Several research topics must be studied to improve our ability to find Earth-like exoplanets. As discussed in this chapter, understanding the relationships between mantle convection and the tectonic regime is required. Most information on that topic would be achieved by comparing Venus and Earth. A better understanding of Venus's interior structure and dynamics would be achieved by dedicated missions to Venus.  The discovery of terrestrial exoplanets also stresses the necessity for a model that combines an atmospheric radiative transfer model with an internal dynamics model. Finally, it is crucial to determine the mass of the small terrestrial exoplanets and to get additional information on their atmospheric composition. Such measurements may be obtained from Earth-based very large telescopes or from space telescopes (see \S6).

\bigskip

\centerline{\textbf{ 4. GIANT PLANETS IN THE SOLAR SYSTEM}}
\bigskip

Much like the Sun is our reference standard for stars, the Solar System's giant planets are our standards for giant planets.  They can be observed in great detail from Earth, and space missions such as \emph{Pioneer}, \emph{Voyager} 1 and 2, \emph{Galileo}, and \emph{Cassini} can provide refined measurements of important quantities, including the planetary gravity field.  \emph{In situ} measurement has also taken place, thanks to the \emph{Galileo Entry Probe}.  Our four giant planets show incredible diversity in physical properties, as perhaps one might expect given the factor of 20 difference in mass between Jupiter, and the much smaller Uranus and Neptune.  Our luck at having four nearby examples of giant planets for study, which could well be a rarity in planetary systems, has allowed us to appreciate the complexity of these planets.

In this section we will first discuss the key observations of our Solar System's giant planets and then our ``classical" views of their structure.  These will be used as jumping off points to understand recent work which in some cases has dramatically revised our understanding of these planets.

\bigskip
\noindent
\textbf{4.1 Classical Inferences for the Solar System Planets}
\medskip

With knowledge of the EOS of hydrogen and helium under high pressure, one can compute a cold-curve ($T$=0) mass-radius relation for a solar H-He mix.  Such curves, over a wide range in mass, were computed by, for instance \ct{Zapolsky69}.  This shows that Jupiter and Saturn are predominantly H/He objects.  One can also compute similar curves for adiabatic models of the planets, where interior temperatures reach $\sim$10,000-20,000 K \cp[e.g.][]{Fortney07a,Baraffe08}.  Both Jupiter and Saturn are found to be smaller and denser than pure H/He adiabatic objects, with Saturn farther from pure H-He composition.  Thus one can tell from the planets' masses and radii alone that they are enhanced in metals compared to the Sun.  Similar calculations for Uranus and Neptune suggest that the planets are predominately composed of metals, with a minority of their mass in the H-He envelope that makes up the visible outer layers.  More sophisticated models can yield additional information, but in the era of exoplanets it is always important to keep in mind that much can be learned from mass and radius alone.

There are actually a reasonably large number of observables beyond mass and radius that can be used to better understand the current structure of giant planets.  These include the rotation rate, equatorial radius, polar radius, temperature at a 1-bar reference pressure, total flux emitted by the planet, total flux scattered by the planet, and the gravity field.  Historically, the oblateness of the planet, and the gravity field, when combined with measured rotation rate, have yielded useful constraints on the planetary density as a function of radius.  The planets rotate, and how they respond to this rotation via their shape, and via a gravity field that differs from a point mass, yields essential information.

The gravity field is generally parametrized by the gravitational moments $J_{\rm n}$, which are the leading coefficients if the external gravitational potential 
is expanded as a sum of Legendre polynomials 
\cp[see e.g. Eqs.(10)-(11) of][]{FortneyBaraffe11}.
%
%
These coefficients, ``the Js," can be measured by observing the acceleration of spacecraft via Doppler shift of their emitted radio signals.  In some cases the coefficients can also be constrained from long term motions of small moons.   A method is then needed to calculate $J_{\rm n}$ based on an interior structure model that obeys all observational constraints.  The state of the art for many decades has been the ``Theory of Figures," as described in full detail in \ct{ZT78}.  At this time the method is accurate enough for  calculations out to $J_{\rm 6}$.  
%
%

The classical view of the planets was well-solidified by around 1980, when the core-accretion model of \ct{Mizuno80} suggested the giant planets needed $\sim$10 \me\ primordial cores in order to form.  Around this same time \ct{Hubbard80b} found that the structure of all four giant planets were consistent with $\sim$10-15 \me\ cores at the current day.  Even very precise knowledge of the gravity is of limited help when directly constraining the mass of the core.  This is because the gravitational moments predominantly probe the outer planetary layers, and the weighting moves closer to the surface with higher order \cp[see Fig.1 of][]{Helled11c}.
For Jupiter and Saturn, the region of the core is not directly probed, while for Uranus and Neptune, one has more leverage on core structure.


Models of the thermal evolution of giant planets aim to understand the flux being emitted by the planets.  There are two separate components, since our relatively cool giant planets are warmed by the Sun.  These components are generally written in terms of corresponding temperatures.  \ttherm\ characterizes the total thermal flux emitted by the planet.  \tint\ characterizes the thermal flux due only to the loss of remnant formation/contraction energy, which is much larger at earlier ages when a planet's interior is hotter.  \teq\ characterizes the component that is due only to absorbed solar flux, which is then re-radiated to space. This component can also be time varying (to a lesser degree) due to changes in the solar luminosity, which are readily understood, and changes to the planetary Bond albedo, which are harder to model.  With these definitions, at any age \ttherm$^4$ = \tint$^4$ + \teq$^4$.  At a very young age \tint$^4$ $>>$ \teq$^4$, while today 
\tint $<<$ \teq  \,and \ttherm $\sim$ \teq.

Cooling calculations yield the planetary \ttherm\ over time, and a correct model would match at least the current \ttherm\ and radius of the planet at 4.5 Gyr.  There can also be other relevant observational constraints.  Cooling models originated with Jupiter in \ct{Graboske75} and within a few years \cp{Hubbard77,Bodenheimer80} it was clear that a Jupiter could have cooled to its present \ttherm\ of 125 K from a hot, high entropy phase, in the age of the Solar System.  However, \ct{SS77b} and \ct{Pollack77} showed that Saturn is currently much more luminous (by $\sim$50\%) than one obtains from a similar simple model.  It has long been suggested that a key to understanding this model ``cooling shortfall'' is in an additional energy source within Saturn due to the differentiation of the planet (see \S 2.3).  \ct{Stevenson75} suggested that, at megabar pressures and temperature below $\sim$10,000 K, H and He could phase separate, leading to helium-rich droplets raining down within a giant planet \cp{SS77b}.  This helium rain would essentially be a change of gravitational potential energy into thermal energy (see \S 2.3).

From the \emph{Galileo Entry Probe} it is known that the atmosphere of Jupiter is modestly depleted in helium, with a mass fraction $Y=0.234 \pm 0.005$ \cp{vonzahn98} compared to the protosolar value of $0.2741 \pm 0.0120$ \cp{Lodders03}, signalling that helium phase separation is underway (see \S 2.3).  Jupiter's atmosphere is also strongly depleted in neon, which seems to be preferentially incorporated in the He-rich phase which sediments out, as shown by theoretical calculations  \cp[e.g][]{Wilson10}.  In Saturn, where He phase separation should be more pronounced, there has been no entry probe, and the helium abundance is much more uncertain.  \ct{CG00} suggest $Y=0.18-0.25$.  Thermal evolution models with He phase separation have been calculated by \ct{Hubbard99} and \ct{FH03} for Saturn, suggesting that if He rains down nearly to the planet's core, then the high planetary \ttherm\ and relatively modest atmospheric $Y$ depletion can be simultaneously matched.

Models of the thermal evolution of Uranus and Neptune have had to be generally more exploratory, since the main components are not known with certainty, and the EOS of the components have not been studied in as much detail.  However, they has long been an indication that both planets are actually under-luminous compared to a simple model with an adiabatic interior \cp{Ubook,Nbook}.  However, with recent advances in the water EOS \cp[see \S 2.2;][]{French09} evolutionary models are being revisited \cp[][see \S 4.5]{Fortney11}.

\bigskip
\noindent
\textbf{4.2 Common Model Assumptions}
\medskip

Modelling the interior structure of giant planets and comparing these models with observations is fraught with degeneracies.  This is because planetary density and gravity field measurements are integrated quantities that sample large fractions of the planetary interior.  With the acknowledgement that some of these assumptions may not hold in the interiors of planets, workers have traditionally forged ahead while making traditional assumptions.  These are a fully convective (and hence adiabatic and isentropic) interior, a central heavy element core that is distinct from the predominantly H-He envelope, a composition in the H-He envelope that is entirely uniform, or one that is broken up into two layers, which may differ in helium or metal content, solid body rotation or rotation on cylinders, and a rotation period for the deep interior that is given by the measured magnetic field rotation rate.  Additionally for Uranus and Neptune, the deep interior is broken into two heavy element layers, meaning a rocky core, a water-rich middle envelope, and predominantly H-He outer envelope.

Therefore most models of giant planets have 3-layer structure.  For Jupiter and Saturn, these three layers from the inside out would be:  layer 1, core material (which could be made of rock and/or water, in a mixture or in two layers); layer 2, an inner envelope made predominantly of liquid metallic hydrogen, which is often enhanced in helium, with a prescribed metal mass fraction; and layer 3, an outer layer of predominantly fluid molecular hydrogen, which is often depleted in helium, with a prescribed metal mass fraction.  The boundary between layers 2 and 3 is often left as a free parameter, but is often around 1-2 Mbar, the approximate transition region for hydrogen to change from molecular to metallic form.  This is also the region where He is expected to be most immiscible, so that placing a He discontinuity here makes some physical sense.  If the giant planets are fully convective, it is not as clear that discontinuity in metals should also occur at this \emph{same} pressure, since mixing could homogenize the metals.  However, the dredge-up of any heavy element core could lead to metal enhancement in the deeper regions.  The values of the mass of any core, and the amount of metals within the H-He envelope are iterated until the calculated $J_n$ values agree with observations.

Within Uranus and Neptune, the three layer structure is manifest in a similar manner.  The inner core is assumed to be rock or rock/iron.  The middle layer is mostly water or some mixture of fluid icy volatiles (like H$_2$O, NH$_3$, CH$_4$).  The outer envelope is predominantly fluid H$_2$ and He.  Pressures in the outer layer are not high enough for liquid metallic hydrogen to form.  However, to match the gravity fields of the planets, often the middle layer must be less dense that an icy mixture (so that H-He is included in the middle layer) and the outer envelope is more dense than H-He, so that an admixture of icy material is included in the H-He layer.  Since metals make up most of their mass, one would certainly be interested for more detailed information, in particular the ice-to-rock ratio in Uranus and Neptune.  However, one can arrange mixtures of rock, water, and H-He in a variety of reasonable configurations, yielding a variety of degenerate solutions.

\bigskip
\noindent
\textbf{4.3 Results from Classic Models}
\medskip

``Classic'' models on the interiors of the giant planets remain relevant for a number of reasons.  Perhaps most importantly the relevant input physics has often been poorly known.  To put it simply, if one does not understand the EOS of hydrogen well, which is the most important component, there is potentially little to be gained by adding additional complications to a structure model.  Even in the era of the advent of more sophisticated models, classic models will still represent an important area of work.  But we are now in the position to more properly evaluate if common assumptions are indeed true.

Models of Jupiter and Saturn in particular aim to constrain the total metal mass fraction within the interior, and what fraction of these metals are found in a heavy element core.  Figure \ref{jsint} summarises the results of modern adiabatic models.  These are models computed by several groups, using different EOSs \cp{Militzer08,Fortney10,Nettelmann13a,Helled13}.  

For Jupiter, the upper left region in Figure \ref{jsint} is from \ct{Militzer08}, who used a first-principles H/He EOS.  The larger region on the bottom right and down is from \ct{Fortney10}, who use a wide range of possible H/He EOS, and also allow for a possible discontinuity in the heavy element enrichment in the H/He envelope. Models connecting these two regions are therefore also plausible, given the uncertainty in the H/He EOS.

For Saturn, the left-side region is from \ct{Helled13}, who considered models with a homogeneous H/He envelope and the SCVH EOS.  The larger region on the right is from \ct{Nettelmann13a} who considered a wider range of H/He EOS, and also allow for a discontinuity in heavy elements in the H/He envelope. \ct{Nettelmann13a}  also find a range of models that essentially encompass the \ct{Helled13} results as a subset.  \ct{Nettelmann13a} models with small cores generally have very large enrichments in heavy elements, up to 30 times solar, in the deep region of the H/He envelope.  Therefore, even with small cores, Saturn remains having appreciable central concentration. 

Note that if Jupiter were precisely of solar composition, it would contain 1.4\% metals by mass \cp{Asplund09}, or 4.5 \me.  For Saturn this would be 1.3 \me.  On the whole planets are substantially enhanced in metals, a factor of $\sim$3 to 8 for Jupiter, and $\sim$12 to 21 for Saturn, compared to the solar composition.
\begin{figure}[h]
 \vspace{-6cm}
  \includegraphics[scale=0.4]{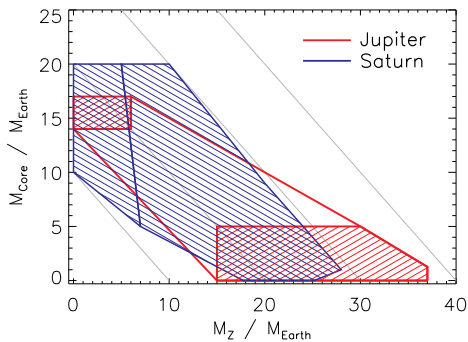}
  \caption{\small Inferred constraints on the interior of Jupiter and Saturn from fully convective models with a distinct core.  The y-axis shows metals in the core, while the x-axis shows metals within the H-He envelope.  Gray lines are constant metal mass. See text (\S 4.3) for explanations of hatched regions and lines.
 } \label{jsint}
 \end{figure}

For Uranus and Neptune, one would like to constrain the ice-to-rock ratio in the interior, as well as the distribution of these metals.  The gravity fields for these planets are less well known.  However, the main obstacle is the significant degeneracy in possible composition.  These issues were reviewed in detail in \ct{Ubook} and \ct{Nbook}.  A mixture of high pressure rock and H-He has an EOS that can closely mimic that of water.  Therefore it is possible to construct acceptable models of the planets \emph{without} water/ice of any kind, if the bulk of the interior is a high pressure rock-H-He mixture.  Also, evidence for a high density rocky core is not particularly strong.  Models with a compressed water-like density throughout most of the deep interior are acceptable.

In Figure \ref{unint} we show calculations from \ct{Nettelmann13b} of current constraints on the interior of Uranus and Neptune.  These models constrain the metallicity of the ``water-rich'' inner envelope with metallicity $Z_2$, and the H-He rich outer envelope, with metallicity $Z_1$.  As discussed in \ct{Nettelmann13b} the gravity field of Uranus is better constrained than that of Neptune, so \emph{within the framework of 3-layer adiabatic models} its structure is better constrained.  In these models the fluid ices are modelled using the water EOS of \ct{French09}.  For both planets, the inner water-rich envelope must have an admixture of a lower density component (here H-He in solar proportions). 

\begin{figure}[h]
 \vspace{-2cm}
 \hspace{1cm}
 \includegraphics[scale=0.3]{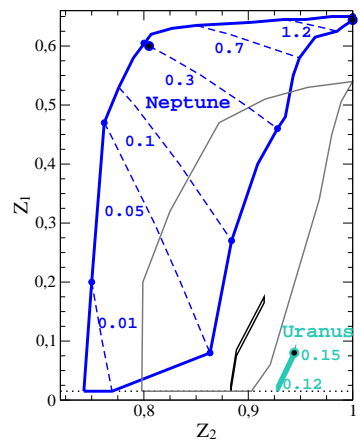}
   \vspace{-1cm}
  \caption{\small Adapted from \ct{Nettelmann13b}.  Heavy element mass fraction in the outer envelope ($Z_1$) and inner envelope ($Z_2$) of Uranus models (black) and Neptune models (grey) as labelled with the Voyager shape and rotation data. The solid lines frame the full set of solutions for each planet.  Dashed lines within the box of Neptune models indicate solutions of same transition pressure between inner and outer envelopes (in [Mbar] as labelled).  
  The dotted line is a guide to the eye for the solar metallicity $Z$ = 0.015.  The colored areas use the modified shape and rotation data for Uranus (green) and Neptune (blue).
}  \label{unint}
 \end{figure}

\bigskip
\noindent
\textbf{4.4 A More Modern View of Jupiter and Saturn}
\medskip

Over the past few years researchers have begun to question in ernest some of the long-held assumptions of giant planet modelling, in particular for the gas giants Jupiter and Saturn.  These assumptions include a fully adiabatic interior and the likelihood that a distinct core exists.  Some 20 years ago \ct{Guillot94a} investigated whether the giant planets may have radiative windows in the H$_2$-dominated outer layers.  A followup study \cp{Guillot04} using modern opacities well-tested again brown dwarf atmospheres found that adiabatic convection did hold.

\ct{Stevenson82a} and \ct{Stevenson85} suggested that there was no sound reason to believe the primordial cores would remain distinct from the overlying H-He envelope, and \ct{Guillot04} put forth a simple model of how these cores could be dredged up by convection over time.  As described in \S 2, with the rise of detailed \emph{ab initio} EOS calculations, it has become feasible to study the miscibility of water, rock, and iron in liquid metallic hydrogen at high pressure.  B.~Militzer and collaborators have shown that each of these components is miscible at the pressures and temperatures relevant to Jupiter's core \cp{Wilson12,Wilson12b,Wahl13}.  In Saturn, much the same story holds, but the lower temperatures may lead rocky materials to stay unmixed.

If core material is able to be mixed into the overlying H-He envelope, the next natural questions are ``Does this occur?" and ``How does it affect the planet's evolution?"  The redistribution of any core material is not a simple problem.  It takes work for the planet to gradually bring up the denser material (be it watery or rocky) and the efficiency of this process is essentially not understood.  One could imagine that a core may be gradually diluted into the overlying H/He mixture.  The composition gradient may then extend over the interior of the planet, from a highly metal-rich central region, with a smaller metal mass fraction with increasing radius.  The process would be an energy sink, since it would alter the gravitational energy of the planet to make the planet less centrally condensed.  Much of this redistribution could have potentially happened quite early on, when the planet's interior was hotter and convection more vigorous.

Only very recently have researchers begun examining how such composition gradients would affect the past, current, and future structure and cooling of giant planets.  \ct{Leconte12,Leconte13}, extending earlier work by \ct{Chabrier07c}, have examined the interior structure and thermal evolution of Jupiter and Saturn with heavy element composition gradients throughout most of the interior of the planets.  They have used results of extensive 3D calculations of energy transport in the ``double-diffusive" regime where both temperature and a dense component may be redistributed \cp{Rosenblum11,Mirouh13}.

\ct{Leconte12,Leconte13} investigate how composition gradients, based on parametrizations of the efficiency of energy transport within the interiors of the planets, change the temperature stratification of the interior.  For Jupiter and Saturn they find interior structure models that are hotter than homogeneous adiabatic models, and therefore they are more heavy-element rich to match the planet's gravity field constraints.  They also have calculated thermal evolution models for Saturn than can match the planet's anomalously high intrinsic flux without the need for helium phase separation.  Composition gradients suppress interior cooling, such that compared to a standard adiabatic cooling model, intrinsic fluxes are lower at young ages, but  are  in turn higher  at old ages.  These new models are important and suggestive.  To ``close the loop" on a modern alternative view of giant planet interiors, the final piece of the puzzle, mentioned above, is an understanding of how planet formation, or rapid core dredge up at young age, could lead to the kinds of composition gradients studied by Leconte \& Chabrier.

\bigskip
\noindent
\textbf{4.5 A More Modern View of Uranus and Neptune}
\medskip

While in the past 20 years Jupiter has been visited by \emph{Galileo} and soon by \emph{Juno}, and \emph{Cassini} has been at Saturn for nearly a decade, Uranus and Neptune were only visited via a single flyby by \emph{Voyager 2}.  Progress in our understanding of these planets has been slowed by a lack of data compared to Jupiter and Saturn, a lower precision on measured quantities, and finally, the inherent complexities of the planets themselves.

Recent progress in understanding Uranus and Neptune have come from a new appreciation of past measurement uncertainties, as well as from the application of new EOS for modelling the interior structure and thermal evolution of the planets.  \ct{Helled11c} confirmed previous results from \ct{Marley95} that the gravity field of the planets do not require having distinct compositional layers within the planets.  \ct{French09} have published a new \emph{ab initio} EOS for water, which has been shown to agree very well with high pressure shock data \cp{Knudson12}.  Thermal evolution models of Uranus and Neptune that use this new EOS can lead to dramatically revised cooling histories.  \ct{Fortney10} and \ct{Fortney11} found that they could match the measured \ttherm\ of Neptune at 4.5 Gyr without having to resort to ad-hoc cooling histories.  Models with fully adiabatic interiors, and 3 distinct layers (the simplest possible model) match the planet's cooling history.  However, evolutionary models for Uranus still show a large difference between the predicted intrinsic flux from the planet, and the lower measured flux.

Most recently, \ct{Kaspi13}, following up on work from \ct{Hubbard91}, have shown that the available gravity field data for Uranus and Neptune strongly constrain the depth to which the visible atmospheric differential rotation penetrates into the interior.  They show that $J_4$ constrains the dynamics to the outermost 0.15\% of the total mass of Uranus and the outermost 0.2\% of the total mass of Neptune.  Therefore the dynamics visible in the atmosphere are confined to a thin layer no more than about 1,000 km deep on both planets.  Similar constraints for Jupiter and Saturn are not yet available, but should be in 2017 from new spacecraft data described below. 

Finally, but quite importantly, there has recently been extensive work on discussion on the true rotation periods of all of the giant planets, whether the measured magnetic field rotation periods are consistent with the measured oblateness measurements of the planets, and implications for interior models \cp{Helled09,Helled10}.

\bigskip
\noindent
\textbf{4.6 Seismology of the giant planets}
\medskip

Seismological data has been essential to our current understanding of the internal structure of the Earth, the Sun, and other stars.  There has also long been interest in measuring the oscillations of the Solar System's giant planets, as such data would at a minimum be complementary, at best significantly more constraining, than gravity field data.  A definitive detection of oscillations of any Jovian planet could in principle serve to accurately determine the core radius and rotation profile of the planet. Since such determinations would remove two important sources of uncertainty surrounding the interior structure, more information could then be gleaned from the traditional interior model constraints. Seismology might also help to constrain more accurately the location of any layer boundaries, such as the transition from molecular to metallic hydrogen or a region of helium enhancement.

After a long history of ambiguous observations, there have recently been two observational breakthroughs.  The first was the detection by \ct{Gaulme11} of Jupiter's oscillations with the SYMPA instrument \cp{Schmider07}, with a measurement of the frequency of maximum amplitude, the mean large frequency spacing between radial harmonics, and the mode maximum amplitude.  In particular, the large frequency spacing of $155.3 \pm 2.2$ $\mu$Hz, was fully consistent with most computed models of Jupiter's interior.

The other avenue into giant planet seismology is ``using Saturn's rings as a seismograph," which was discussed in detail by \ct{Marley91}.  Going back to \ct{Rosen91}, it was found that there were spiral structures in Saturn's rings seen with \emph{Voyager 2} that could not be explained by any satellite resonances, but could reasonably be explained due to perturbations on the rings from Saturn free oscillation \cp{Marley93}.  Very recently \ct{Hedman13} used superior \emph{Cassini} data to definitely confirm the presence of features in the rings at locations suggested by \ct{Marley91}, and ruled out moon-based resonances based on the wave speeds.  While the orbital separations of the ring features were as expected, the unexpected feature from the \ct{Hedman13} analysis was that they identified multiple waves with the same number of arms and very similar pattern speeds.  This indicates that multiple $m = 3$ and $m = 2$ sectoral ($l = m$) modes may exist within the planet.  This peculiar feature was not anticipated from any Saturn model.  The methods of modern helioseismology, now brought to bear on giant planets, may allow for important advances \cp[e.g.][]{Jackiewicz12}.


\bigskip
\noindent
\textbf{4.7 \emph{Juno} at Jupiter and Future Space Missions}
\medskip

In 2016 the NASA \emph{Juno} spacecraft will enter a low periapse polar orbit around Jupiter.  Many of the major science goals for \emph{Juno} involve answering major questions about the interior of the planet.  First, the mission will accurately map the gravitational field of the planet at high accuracy out to very high order, perhaps $J_{12}$.  There will also be sensitivity to non-zero odd $J_n$ values, such that deviations from hydrostatic equilibrium can be constrained.  The degree to which the outer layers of the planet rotate differentially, on cylinders, or via solid body rotation will be strongly constrained.  Tides raised on the planet by Io may be detectable, which would lead to a measurement of the $k_2$ tidal Love number \cp{Hubbard99}.  The Lense-Thirring Effect is probably observable and will enable a direct measurement of the planet's rotational angular momentum, which could then strongly constrain the planet's moment of inertia \cp{Helled11b}.

The magnetic field of the planet will also be extensively mapped.  While in principle there should be tremendous synergy between interior modelling to better understand composition and structure, and interior models to understand convection and the interior dynamo, in practice these communities do not have enough cross-talk.  With outstanding data sets on both gravity and magnetic field, this will hopefully improve.  Finally, \emph{Juno} will have a microwave radiometer that will allow spectroscopy of the planet's thermal emission down to a depth of $\sim100$ bars.  This depth is important, since the \emph{Galileo Entry Probe} only reached 22 bars, which apparently was not deep enough to reach below the level where water vapor is partially condense into clouds, such that the measured water abundance was only a lower limit.  With \emph{Juno} the mixing ratios of H$_2$O and NH$_3$ should both be measured, which are two of most abundant molecules (along with CH$_4$) in the H-He envelope.  With a refined understanding of the mass fraction of heavy elements within the H-He envelope, the total mass of metals within the planet will be better constrained. 

In 2017, the \emph{Cassini Solstice} extended mission at Saturn is scheduled to end.  The current plan for the final stages of the \emph{Cassini} spacecraft is to bring it into a some-what \emph{Juno}-like low periapse polar orbit, so that precision mapping of the planet's gravity and magnetic fields can be undertaken, in gravity up to $J_{10}$.  Much of the same work on understanding differential rotation inside the planet will be enabled.  Again in analogy to \emph{Juno} at Jupiter, it has been suggested by \ct{Helled11a} that the Lense-Thirring Effect could be observed, which would further constrain the current interior structure.   Eventually the craft will be manoeuvred down into Saturn's atmosphere, where it will break up.  It may be possible for \emph{in situ} sampling of Saturn's atmospheric mixing ratios by the mass spectrometer before breakup.

Given that precise gravity field data will be available to high order, new methods are now being developed to allow for highly accurate calculation of the gravity field from internal structure models.  These new methods, which deviate from the Theory of Figures developed in the 1970s \cp{ZT78}, are now just being published \cp[e.g.][]{Hubbard12,Kong13,Hubbard13}, but they are an exciting advance.  The generalization of these models beyond merely 1D structure models should also be expected.

There are currently no plans for missions to the Uranus or Neptune systems.  Given that Neptune-class planets significantly outnumber true gas giants in exoplanetary systems, it is essential to better understand our lower-mass giant planets.  Proposed European M-class (medium class) missions that would venture to either Uranus or Neptune in the mid 2030s are Outer Solar System (OSS) to Neptune \cp{Christophe12} and Uranus Pathfinder \cp{Arridge12}.

%

\bigskip

\centerline{\textbf{ 5. EXTRASOLAR GIANT PLANETS}}

\bigskip
\noindent
\textbf{5.1 Status of current models}
\medskip

There are two key physical ingredients which characterise state-of-the-art models for internal structure and evolution of exoplanets. The first one is the use of  modern EOSs, as described in \S 2, which allow a variation of  mixtures and amounts of heavy elements to account for the wide variety of  bulk compositions observed for transiting planets. The second one is the use of accurate atmospheric boundary conditions, taking into account up-to-date opacities and irradiation from the parent star for close-in planets (see contribution by Madhusudhan et al., this book).

\medskip\noindent
{\it Heavy material enrichment} 
\smallskip

As mentioned in \S 4.1,  the impact of heavy material enrichment on planetary radius dates back to the pioneering work by \cite{Zapolsky69}, who studied the mass-radius relationship for objects composed of zero-temperature single element. If the energy transport in the interior is due to efficient convection, the internal temperature profile is quasi-adiabatic. The mechanical structure and thermal internal profile thus strongly depend on the EOS of its chemical constituents. 

Figure \ref{fig_mr} shows a spread in radius for a given planetary mass, which is partly due to diversity in the bulk chemical composition of planets. Such diversity is expected from our current understanding of giant planet formation processes (see e.g contribution by Helled et al., this book). Additional physical processes, required to explain abnormally large radii observed for a significant fraction of close-in exoplanets and discussed in \S 5.2, are also certainly responsible for a spread in radius. While an increase of the interior heavy element content in a giant planet contributes to a decrease of its radius (see Fig. \ref{fig_mr}), the additional physical processes afore-mentioned work in reverse and contribute to increasing its radius. Since the mechanisms responsible for inflation of close-in exoplanets have not yet been firmly identified, 
inferring their bulk composition from the knowledge of their mass and radius is 
still highly uncertain. As discussed recently in \cite{Miller11}, a relationship between stellar metallicity and planetary heavy elements seems to exist, as originally suggested by \cite{Guillot06}.
Such relationship, if confirmed and determined with some confidence, could be used to infer the amount of heavy material in a given inflated close-in planet, based only on the parent star metallicity and planet mass. This would provide a powerful tool to determine the amount of additional energy needed to explain a planet's inflated radius. 

Currently, the uncertainty resulting from the lack of knowledge of the inflation mechanisms (see \S 5.2) adds to the uncertainty in current EOS used to model planetary interiors. The most widely used EOSs to describe the thermodynamic properties of exoplanet's interiors are the Saumon-Chabrier-Van Horn EOS for H/He mixtures \citep{SCVH}, and ANEOS or SESAME  for heavy elements, usually water, rock or iron (see \S 2). These EOS have been used to construct models covering a wide range of planetary masses including different amounts and compositions of heavy material \citep{Guillot06, Burrows07, Fortney07a, Baraffe08}. A detailed analysis of the main uncertainties in planetary models which may result from the choice of the EOSs and the amount and distribution of heavy material within the planet has been conducted by \cite{Baraffe08}. 
This work shows the sensitivity of the radius to the choice of the EOS (i.e ANEOS versus SESAME) and to the distribution of heavy material within the planet. 
The largest difference between the various EOS models, reaching up to $\sim 40$-60$\%$ in $P(\rho)$ and $\sim 10$-15$\%$ on the entropy $S(P,T)$,
occurs in the $T\sim10^3$-$10^4$ K, $P\sim 0.01$-$1$ Mbar interpolated region, the typical
domain of Neptune-like planets. 
The radius of a 20 $\mearth$ planet having 50\% heavy element enrichment can vary by  25\% depending on whether the heavy material is in a central core or distributed over the whole planet. Such high sensitivity of the radius is due to differences in entropy predicted by the different heavy material EOSs used and by different impacts of the entropy of heavy material whether it is in a core or mixed with H/He in the envelope \cp{Baraffe08}.
 Improved EOS as described in \S 2 are now starting to be used to model the Solar System giant planets (see \S 4) and specific exoplanets \citep{Nettelmann10, Nettelmann11}. A preliminary revision of the mass-radius relationship for exoplanets based on ab initio EOS for H and He shows deviation compared to models based on the Saumon-Chabrier-Van Horn EOS \citep{Militzer13}. Extensive work is now required to systematically investigate the effects of new ab initio EOSs on the structure and evolution of exoplanets, promising some exciting results in a near future. It will be of particular interest to know whether improved EOSs confirm or not the sensitivity of the radius to the distribution of heavy material within a planet.


\medskip\noindent
{\it Irradiation effects} 
\smallskip

Because of the high proximity of many newly discovered exoplanets from their parent star, due to current detection technics, the heating from the incident stellar flux of the planet atmosphere needs to be accounted for. The incident flux can be defined by: 

 \begin{equation}
F_{inc} = \alpha\left(\frac{R_{\star}}{r(t)}\right)^2 \sigma T_{\star}^4,  
\end{equation}

where $R_{\star}$
and $T_{\star}$ are the host star radius and effective temperature respectively, and $\alpha \in [1/4,1]$
is a scaling factor used to crudely account for day-to-night energy redistribution. 
The latter represents a proxy for atmospheric circulation, derived by considering whether the absorbed stellar radiation is redistributed evenly throughout the planet's atmosphere ($\alpha = 1/4$), e.g due to strong winds rapidly redistributing the heat,  or whether the stellar radiation is redistributed and reemitted only over the dayside ($\alpha = 1/2$). Since exoplanets are found with a wide range of eccentricities, the general (time-dependent) planet-star separation is given by:

\begin{equation}
r(t) = \frac{a(1 - e^2)}{1 + e\cos\theta(t)},
\end{equation}

where $a$ is the semi-major axis, $e$ is the eccentricity, and
$\theta(t)$ is the angle swept out by the planet during an orbit. Like for Solar System giant planets
(see \S 4.2),
an useful characteristic atmospheric temperature for irradiated
exoplanets  is \ttherm\, the total thermal flux emitted by the planet, given by:

\begin{equation}
\sigma T_{\rm therm}^4 = \sigma T_{\rm int}^4 + (1 - A)F_{\rm inc},
\end{equation}

with  $T_{\rm int}$ the internal effective temperature, which is a measure of the energy contribution from the planet interior (see \S 4.2), and $A$ the bond albedo. The second right-hand term defines the equilibrium temperature $\sigma T_{\rm eq}^4 = (1 - A)F_{\rm inc}$ mentioned in \S4 for our giant planets.


For planets of a few Gyr, $T_{\rm int} << T_{\rm eq}$ and $T_{\rm therm} \sim T_{\rm eq}$. In this case, as already stressed in \cite{Barman01}, an irradiated atmospheric structure characterised by  $T_{\rm eq}$ can differ significantly from a non-irradiated structure at the same effective temperature $\te = T_{\rm eq}$. This point \cp[see also][]{Seager98, Guillot02} emphasizes the fact that adopting outer boundary conditions, for evolutionary calculations, from atmospheric profiles of non irradiated models with $\te = T_{\rm eq}$ is incorrect and yields erroneous evolutionary properties for irradiated objects.
More correct outer boundary conditions for evolutionary models are based on sophisticated 1D static atmosphere models including state-of-the-art chemical and radiative transfer calculations, with the impinging radiation field explicitly included in the solution of the radiative transfer equation and in the computation of the atmospheric structure  \citep{Barman01, Sudarsky03, Barman07, Seager05, Fortney06, Burrows08}. Irradiation produces an extended radiative zone  in the planet outer layers and below the photosphere, pushing to deeper layers (i.e larger pressure) the top of the convective envelope compared to the non-irradiated counterpart. 
The main effect of irradiation on planet evolution is to slow down the contraction and produce a radius slightly larger than the non irradiated counterpart at same age. For known planetary systems, the effect is rather modest for hot Jupiters, with an increase in radius of less than $\sim$ 10\% \cp{Baraffe03}. For lower mass planets, however, the effect can be significant,
as illustrated by the red curves in Fig. \ref{fig_mr}. In a recent work, \cite{Batygin13} highlights the fact that the radius of an irradiated  super-Earth planet may exceed that of its isolated counterpart by as much as a factor $\sim$ 2.
Interestingly enough, this brings the radius of such a planet well into the characteristic range of giant planets. 

\medskip\noindent
{\it Atmospheric Evaporation} 
\smallskip

It is now well admitted that atmospheric escape may play an important role in irradiated, close-in planets and shape their faces. Planet evaporation can take place through a variety of mechanisms: non-thermal, thermal Jean's and hydrodynamic escapes. Hydrodynamic evaporation or "blow-off" may be experienced by hydrogen-rich upper atmospheres of close-in giants orbiting at less than 0.1 AU and heated up to temperatures of more than 10 000K by X-rays and UV radiation from the parent star \citep{Lammer03, Yelle04}. This mechanism has received most attention in the exoplanet literature since it is  the only one which may produce high enough mass loss to affect the structure and evolution of close-in planets. Originally, models of XUV-driven mass loss were first developed to study water loss from early Venus \citep{Hunten82, Kasting83} and hydrogen loss from the early Earth e.g \cite{Watson81}. These kinds of models were extended  and applied to the study of hot Jupiters 
 \citep[e.g][]{Lammer03, Yelle04, Tian05, Murray09, Koskinen10, Ehrenreich11, Owen12}. 
Those objects remain the best studied cases for atmospheric erosion of exoplanets.  Mass loss for hot Jupiters typically occurs at a rate $\dot M \sim 10^{10} - 10^{11} g s^{-1}$. HD 209458 b is the first transiting planet for which HI absorption at or above the planet's Roche lobe was observed, suggesting that hydrogen is escaping from the planet \citep{Vidal03}. Further detection of oxygen and carbon in the extended upper atmosphere of this planet confirmed the hydrodynamical escape scenario, or "blow-off" of the atmosphere, where the hydrodynamical flow of escaping HI is able to drag other species like carbon and oxygen \citep{Vidal04}. Despite many debates, at the time of these observations, regarding their interpretation, there is now a consensus that HD 209458 b has an extended exosphere and is losing mass, though details are still not well understood  \cp[see e.g][]{Linsky10}. 

A common approach to estimate the mass-loss rate and study its effect on planet's evolution is to assume that some fixed fraction of the incident stellar XUV energy is converted into heat, which works on the atmosphere to remove mass. This is the so-called "energy limited" approximation \citep{Watson81} which provides a simple analytic description of the mass loss rates \citep{Erkaev07}:

\begin{equation}
\dot M = { \epsilon \pi R^3_{\rm P} F_{\rm XUV} \over G M_{\rm P} K_{\rm tide}}
\label{eqnevap}
\end{equation}

where $F_{\rm XUV}$ is the stellar XUV flux,  $M_{\rm P}$ and $R_{\rm P}$ are  the planet mass and radius respectively, $\epsilon$ is an efficiency factor ($< 1$) which represents the fraction of the incident XUV flux that is converted into work and $K_{\rm tide}$ is a correction factor that accounts for the fact that mass only needs to reach the Roche radius to escape  \cp[see details in][]{Erkaev07}. 

One main uncertainty stems from the determination of the efficiency factor $\epsilon$. 
Estimates from observations of the mass loss rate for HD 209458 b suggest $\dot M \sim 4 \times 10^{10} g s^{-1}$ e.g \cite{Yelle08}, which corresponds to an efficiency factor $\epsilon \sim 0.4$. \cite{Lammer09} suggest that $\epsilon$ is probably less than $\sim 0.6$ for hot Jupiters, but should depend, among other things, on the star's activity and the planet atmospheric composition. An analysis of atmospheric escape from Venus suggests that  $\epsilon = 0.15$ may be a more appropriate value \citep{Chassefiere96}. Finally, the value of $\epsilon$ is certainly not constant over the lifetime of the planet \citep{Owen13}. Despite these uncertainties and difficulties, Eq. (\ref{eqnevap}) certainly provides a reasonable estimate of evaporation of planetary masses by XUV flux and has been often used to study the effect of evaporation on the evolution of exoplanets for a variety of planetary masses \citep{Baraffe04, 
Hubbard07, Jackson10, Valencia10}

The order of magnitude for the mass loss rate estimated for HD 209458 b is confirmed with recent observations for HD 189733 b, another transiting hot Jupiter,  for which atmospheric evaporation has also been detected \citep{Lecavelier10, Bourrier13}.
With such rates, only a few \% of a hot Jupiter mass is lost over its lifetime \citep{Yelle06, Garcia07, Murray09}. This erosion hardly affects a hot Jupiter's internal structure and its evolution, even at very young ages when the high-energy flux of the parent star is expected to be quite large. 
Indeed, \cite{Murray09} demonstrated that, in the case of UV evaporation of hot Jupiters, the energy limited approximation is only valid at low fluxes. At high fluxes, the mass loss process is controlled by ionization and recombination balance, whith photoionization heating balanced by Ly$_\alpha$ cooling.  Mass loss under those conditions is "radiation/recombination-limited" with rates that scale as $\dot M \propto F_{UV}^{1/2}$, instead of $\dot M \propto F_{UV}$ in the energy-limited case, because the input UV power is largely lost to cooling radiation. 

If evaporation is not expected to play a key role on the evolution of giant exoplanets, the story may be different for lower mass exoplanets. Increasing efforts are now devoted to the study of this process for super-Earthes and rocky planets. Because of their lower escape velocities, lower mass planets should have their atmospheres more significantly sculpted by evaporation, like the super-Earthes CoRoT-7 b \citep{Jackson10} and  Kepler-11b  \citep{Lopez12} which could be stripped entirely by UV evaporation. As highlighted by  \cite{Jackson10}, for those low mass planets which may be strongly affected by evaporation and may not have  formed in-situ, the coupling between tidal evolution, affecting the orbital parameters and thus the evaporation rate, and mass loss, affecting the planetary mass and thus the tidal evolution, needs to be considered in order to have a consistent description of their evolution. If this is indeed the case,  evolutionary calculations for these kinds of planets become an extremely complicated and uncertain game.
Models are now also being developed for rocky planets with sub-Earth masses.
As recently suggested by \cite{Perez-Becker13}, even the optical photons from the parent star could vaporize close-in, very low mass rocky planets. Evaporation is thus important for the interpretation of 
observed mass distributions of low mass close-in planets, as provided by Kepler. Improvement in the modelling of evaporation can be done by treating the thermosphere as an integral part of the whole atmosphere, rather than a separate entity as done in many earlier studies, including detailed photochemistry of heavy atoms and ions and more realistic description of heating efficiencies \citep[e.g][]{Koskinen13}. Further developments are expected in the future to better understand this key process which  can strongly affect the original mass of detected close-in, low mass planets and thus the understanding of their formation process.

\medskip\noindent
{\it Deuterium burning planets} 
\smallskip

The IAU definitions for a planet and a brown dwarf, based on the deuterium burning minimum mass, is admittedly practical from an observational point of view but is arbitrary with respect to the formation process.  Characterising these two families of objects by their formation processes rather than using the IAU definition is certainly more natural and physical (see chapter by Chabrier et al., this book).
If planets with a massive core can form above the
aforementioned deuterium burning minimum mass, a key question is to determine whether or not the presence of the core can prevent deuterium burning to occur in the deepest layers  of the H/He envelope. The question is  relevant, since it cannot be excluded that objects forming in a protoplanetary disk via core accretion grow beyond the minimum mass for deuterium burning, e.g 12 $\mjup$ \citep{Kley06}. 
The answer to this question was first provided by \cite{Baraffe08} who showed that in a 25 $\mjup$ planet with a 100 $\mearth$ core, deuterium-fusion ignition does occur in the layers
 above the core and deuterium is completely depleted in the convective H/He envelope after $\sim$ 10 Myr, independently of the composition of the core material (water or rock). 
 A follow-up study has confirmed these results combining core accretion formation models with deuterium burning calculations \citep{Molliere12}.
 These results highlight the utter confusion provided by a definition of a planet based on the
deuterium-burning limit (see chapter by Chabrier et al.).

\bigskip
\noindent
\textbf{5.2 Inflated exoplanets}
\medskip

One of the most intriguing problems in exoplanet physics is the anomalously
large radii of a significant fraction of close-in gas  planets. They show radii larger than predicted by standard models of giant planet cooling and contraction. 
Despite numerous theoretical studies, starting with the work of \cite{Bodenheimer01}, and the suggestion of many possible mechanisms \cp[see the reviews by][]{Baraffe10, Fortney10}, no universal mechanism seems to fully account for the observed anomalies. This implies that either we are still missing some important physics in planetary interiors, or that the solution does no reside in an unique mechanism but in a combination of various processes as described below. 

One can classify the proposed mechanisms into three categories, as suggested by  \cite{Weiss13}: incident stellar flux-driven mechanisms, tidal mechanisms and delayed contraction. This classification is  meaningful in light of increasing evidences for a correlation between the incident stellar flux and the radius anomaly \citep{Laughlin11, Weiss13}. 
Additionally, increasing statistics, thanks in particular to Kepler data, reveal a lack of inflated radii for giant planets receiving modest stellar irradiation \citep{Miller11, Demory11}. Based on a sample of 70 Kepler planetary candidates,  \cite{Demory11} find that below an incident flux of about $2 \times 10^8$ erg s$^{-1}$ cm$^{-2}$, which represents a subsample of $\sim$ 30 candidates, the radii is independent of the stellar incident flux. 
These findings motivate further search for correlations between planetary radius, mass and incident flux.
Additional future data, from Kepler and other surveys, should allow an extension of current studies to planets with longer orbits.   Despite these trends, it is worth discussing mechanisms falling into the three afore-mentioned categories. Indeed, even if some of them cannot explain the aforementioned observed correlations, they may still be important physical processes playing a role at any orbital distance or stellar flux, and thus belonging to the modern theory of planetary structures. We will briefly mention below these mechanisms and their current status as viable explanation for the radius anomaly.

\medskip\noindent
{\it Incident stellar flux-driven mechanisms} 
\smallskip

An attractive scenario which falls into this category is the idea based on atmospheric circulation by \cite{Showman02}.  A close-in, tidally locked planet
receives constantly the stellar flux on the same hemisphere, producing strong temperature contrasts between its day- and night-sides.
The resulting fast winds may produce a heating mechanism
in the deep interior of the planet, slowing down its evolution and yielding a larger than expected radius. This suggestion was based on
 numerical simulations of atmospheric circulation  by \cite{Showman02}, which produce a downward flux of kinetic energy of about $\sim$ 1\% of the absorbed stellar flux. This
 heat flux is supposed to dissipate in the deep interior, producing an extra source of energy during the planet's evolution. The validity of this scenario, however, is still debated and the dissipation mechanism still needs to be found. The substantial transport of kinetic energy found in the simulations of \cite{Showman02} strongly depends on the details of atmospheric circulation models and has not been confirmed by further
 simulations. Important efforts are now devoted to the 
 development of sophisticated 3D dynamical simulations of strongly  irradiated atmospheres  \cp{Showman11}. Most  numerical studies, however, are based on 
 Global Circulation Models (GCMs) developed for the study of Earth weather and climate. They
 usually solve the so-called 'primitive' equations of meteorology,
involving the approximation of vertical hydrostatic
equilibrium, and a 'shallow atmosphere'  \citep[see e.g][]{Vallis06}. These approximations 
may  provide an inaccurate description of the interaction between the upper and deeper 
atmosphere, crucial to test the Showan \& Guillot idea. Developments of state-of-the-art GCM models which solve the full set
of hydrodynamical equations, without any of the previous approximations, and coupled to sophisticated radiative transfer schemes may help solving this problem in a near future \citep{Dobbs12, Mayne13}.

Another mechanism which receives more and more attention is ohmic dissipation. 
The idea is that atmospheric winds blowing across the planetary magnetic field produce currents penetrating the interior and giving rise to ohmic heating in the deeper layers, which could affect the radius evolution of the planet  \citep{Batygin10}. At a similar time, \cite{ Perna10a, Perna10} also suggested the importance of magnetic drag on the dynamics of the atmospheric flows, finding that a significant amount of energy could be dissipated by ohmic heating at depths. This results in energy deposition in the planet given by the ohmic dissipation per unit mass:

\begin{equation}
\dot E = {J^2 \over \rho \sigma}
\label{eohmic}
\end{equation}
with $J$ the current given by Ohm's law and $\sigma$ the electrical conductivity of the gas \citep{liu08, Batygin10}. At temperatures characteristic of hot Jupiters, the primary source of electrons in their atmosphere stems from thermally ionized alkali metals with low first ionisation potentials, e.g Na, Al and K, with K probably playing a dominant role \citep{Batygin10, Perna10}.
The mechanism has been studied based on approximate models for the atmospheric circulation \citep{Batygin10} or using more complex models \citep{Perna10, Rauscher12, Rauscher13}. Ohmic dissipation has been incorporated in calculations of planet thermal evolution, in order to study its effect on the structure and evolution of hot Jupiters \citep{Perna10, Batygin11, Huang12, Wu13}. Most studies 
 suggest that for magnetic field strength greater than $B=3-10$ G, this dissipation can produce enough heat in the planetary interior to slow down its cooling sufficiently to explain the inflated radii of hot Jupiters \citep{Perna10, Batygin11, Rauscher13, Wu13}. However, a recent study by \cite{Huang12} disagrees with these conclusions and finds it difficult for ohmic heating to explain the large radii of hot Jupiters with large masses and large equilibrium temperature (i.e large amount of stellar irradiation received). The discrepancy between their results and previous studies stems from the fact that they do not assume a fixed heating efficiency, namely the fraction of the irradiation going into ohmic power, as in \cite{Batygin11} and  \cite {Wu13}. Instead, \cite{Huang12} use a wind zone model to set the induced magnetic field in the wind zone and the magnitude of the heating.
 
Many uncertainties are inherent to current studies, see e.g  \cite{Rauscher13}, with most calculations of the magnetic drag and ohmic heating based on a formalism that assumes axisymmetry in the flow structure and atmospheric resistivities  \citep{liu08}, which are not realised in highly irradiated planet atmospheres. Simple  geometry for the planetary magnetic field, e.g aligned dipole field, are also usually assumed. Variation of the magnetic geometry, e.g assuming that the planet's magnetic axis is misaligned from its axis of rotation or that the field is multipolar, which may be the case in reality, could have important impacts on the atmospheric circulation. The magnetic drag is usually applied to the zonal flow, although the meridional flow may also experience drag. Another uncertainty stems form the electrical conductivity $\sigma$, which affects the amount of ohmic power, indirectly through its role to determine the current  in the low density wind region and directly since  the ohmic dissipation is proportional to $1/\sigma$ (see Eq. (\ref{eohmic})). Finally, 
the actual wind speed in the layers which may experience the Lorentz drag depends on how deep the wind zone extends, which depends on how well the weather layer couples dynamically to the deeper atmosphere \citep{Wu13}, a process currently not well understood and modelled, as previously discussed.  The final unknown in this process is the strength of hot Jupiter magnetic fields. Based on scaling law arguments, their magnetic field can range between a few and tens of Gauss \citep{Reiners10}.  Direct measure of the planet magnetic field seems difficult since its signal is expected to be buried in the stellar signal. Strategies in the future need to be developed  \citep[see discussion in][]{Rauscher13} to provide a better estimate of these planet magnetic fields and consequently of the significance of ohmic dissipation.
There are thus still many issues and open questions regarding the impact of ohmic dissipation. At the time of this writing, it cannot be concluded whether this mechanism is the universal explanation for hot Jupiter inflated radii or not. 

Two other stellar flux dependent mechanisms have been suggested, namely the thermal tides \citep{Arras10} and the mechanical greenhouse \citep{Youdin10}. They have not received as much attention as the previous mechanisms.  It has yet to be demonstrated that they could explain all of the observed anomalies and we thus 
invite the reader to check the original papers for details.

\medskip\noindent
{\it Tidal mechanisms} 
\smallskip

Tidal dissipation in a giant planet's interior, producing heating that would stop or slow down the planet's contraction, was among the first mechanisms proposed to explain the large observed radius of HD 209458 b \citep{Bodenheimer01}. 
The idea was that an unseen planetary companion would force non-zero orbital eccentricity, which would be tidally damped. The idea was ruled out for HD 209458 b, and for other planets, for which further observations indicated an eccentricity of zero. The idea of tidal dissipation  also
evolved with the discovery of more and more inflated planets, indicating the need for a more general mechanism than the accidental presence of a low mass companion. Since then, this mechanism has received a lot of attention, see \cite{Fortney10, Baraffe10} and references therein. 
With the increasing number of studies gaining in complexity and level of details, the excitement has slowly but surely subsided. This mechanism should certainly be very important for some systems, but the emerging view now is that it cannot be the universal radius inflation mechanism. Radius inflation by tidal heating is, indeed,  a short-lived phenomenon, whereas the average observed planetary system age is several Gyr. As acknowledged  by \cite{Weiss13}, this view was advanced in particular  by \cite{Leconte10} who used the most detailed tidal evolution equations. This paper demonstrates that
quasi-circular approximation, with tidal equations truncated at the order $e^2$ in eccentricity as usually assumed in tidal calculations of transiting planets, is not valid for exoplanetary systems that have, or were born with, even modest value of the eccentricity ($e \simgr 0.2$). In particular, it is shown that truncating the tidal equations at 2nd order in eccentricity can overestimate the characteristic timescales of the various orbital parameters by several orders of magnitudes. Based on the complete tidal equations, \cite{Leconte10} find that tidal heating can explain moderately inflated planets, but cannot reproduce both the orbital parameters and radius of e.g HD 209458 b (see Fig. \ref{fig_hd}) or the radius anomaly of very inflated planets like WASP-12 b which has a radius of $\sim$ 1.8 $\rjup$ \citep{Hebb09}. If tidal dissipation is likely not the universal mechanism responsible for the radius anomaly,
it is still an important mechanism to account for since it impacts the evolution of the orbital properties of planetary systems, playing a role in shaping the distribution of orbits of close-in exoplanets and thus bearing consequences on our understanding of planet formation, migration and planet-disk interaction \citep{Jackson08a}. Orbits of observed hot Jupiters may still be decaying and end up collapsing with their host star, as suggested by \cite{Levrard09}, who show the lack of tidal equilibrium states for many transiting planets, implying that the orbital and rotational parameters evolve over the lifetime of the system.

\begin{figure*}
\vspace{-0.5cm}
 \epsscale{1.}
  \includegraphics[scale=1.4]{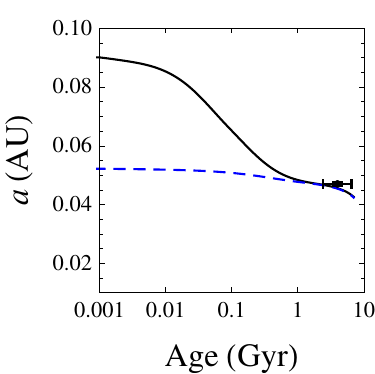}
  \includegraphics[scale=1.4]{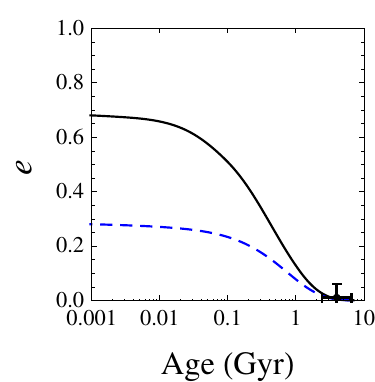}
  \includegraphics[scale=1.4]{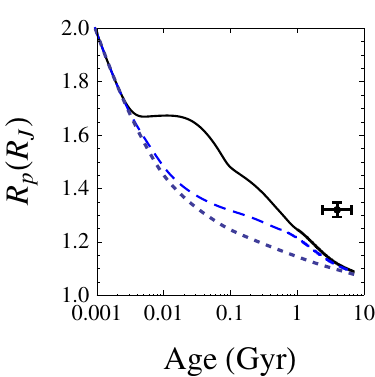}
 \caption{\small Tidal/internal evolution of HD 209458 b with different initial conditions (solid and dashed) computed with complete tidal equations \citep[from][]{Leconte10}. HD 209458 b is a 0.657 $\mjup$ planet orbiting
a 1.01 $\msun$ star (observed parameters with error bars indicated as solid dots in the panels). 
These runs assume for the tidal quality factor $Q = 10^6$ for both the planet and the star  \citep[see][for details and definitions]{Leconte10}. Evolution of the orbital distance (left panel), eccentricity (middle panel) and radius (right panel) are shown.  In the right panel, the radius of an isolated planet (no tidal heating) is also shown for comparison (dotted curve).}
\label{fig_hd}
\vspace{-0.5cm}
 \end{figure*}
 
\medskip\noindent
{\it Delayed contraction} 
\smallskip

The final class of mechanisms that may contribute to larger radii  comprises atmospheric enhanced opacities \citep{Burrows07} and reduced interior heat transport \citep{Chabrier07c}. Regarding the first scenario, \citep{Burrows07} suggested that atmospheres with enhanced opacities would slow down the contraction and cooling of planetary interiors, yielding larger radii at a given age. Enhanced atmospheric metallicities or alteration of atmospheric properties (e.g. photochemical processes driven by strong irradiation, cloud formation) could provide the source of opacity enhancement. This effect may play a role in some transiting planets and affect their evolution, but cannot explain the most inflated transiting planets  \citep[see discussion in][]{Baraffe10}. Also, as noted by \cite{Guillot08},  an overall increase of the planet metallicity to account for the opacity enhancement requires also an increase of the molecular weight in the planet interior which would negate the high opacity effect. 

Finally, following the idea of \cite{Stevenson85}, \cite{Chabrier07c} suggested that heavy element gradients  in planet interior could decrease heat transport efficiency, slowing down its cooling and contraction, and providing an explanation to hot Jupiter inflated radii.  As previously invoked for Solar System giants (\S 4.4), if a vertical gradient of heavy elements is present, resulting from e.g. planetesimal accretion during the formation process or central core erosion, the resulting mean molecular weight gradient can prevent the development of large-scale convection. The combination of heat and chemical species diffusion can drive a hydrodynamical instability and yield so-called 'double diffusive' or 'layered' convection, also known as 'semiconvection' in the stellar community. Because this process does not depend on the stellar incident flux, it is unlikely to be the universal solution to the radius anomaly. But it could be a common property of giant planet interiors, affecting planets at any orbital distance and even our Solar System planets, as mentioned in \S4.4. The work of \cite{Leconte13}, showing that layered convection could explain Saturn's luminosity anomaly (\S4.4),  suggests that the cooling history of giant planets in general 
based on the standard assumption of adiabatic homogeneous structure could be more complex and may need to be revised. The revival of this process for planetary interiors provides a new incentive for the development of  3D numerical simulations \citep{Rosenblum11, Wood13, Zaussinger13a},
which promise advancement in the understanding of semiconvection effects, a long-standing problem in stellar and planetary evolution. 


\bigskip

\centerline{\textbf{ 6. CONCLUSION}}
\bigskip

There has been substantial progress done since the previous Protostar \& Planet meeting in 2005 on the understanding of planetary structures. 
Major advancements  were performed regarding the development of first principle numerical methods to study equations of state for hydrogen, helium and various heavy materials relevant for planet interiors. Ongoing and future high-pressure experiments in various national laboratories (Laser Mega-Joule laser projects in France; Livermore and Sandia in the US) promise substantial progress on this front as well. The perspective to significantly reduce current uncertainties of EOSs in a near future, with such combination of theoretical and experimental works, will allow a significant progress of planetary structure modelling. 

Regarding our Solar System planets, it could be said that the open questions and uncertainties in modelling  giant planets are the same ones from about 30 years ago, since a classic review of \ct{Stevenson82a}.  While for the most part this is true, we are now in an excellent position to make real progress on issues that have been around for quite some time.  In particular the questions of
the overall metal-enrichment of our giant planets, how are these metals distributed and whether they are predominantly originally made from rocky or icy material, will soon be better addressed with the afore-mentioned progress in EOSs and the future space missions to Jupiter and Saturn. Same progress is also expected on inferring the rotation state of Jupiter and Saturn.
To what extent are the interiors of our giant planets, and more generally extrasolar gaseous planets, adiabatic and freely convecting is another long standing problem in the field. Substantial activities to develop 3D hydrodynamical numerical simulations to tackle this problem will shed  important lights on the relevance of semiconvection  in planetary interiors.  

All those questions are also relevant to the modelling of exoplanets.  At the time of this writing, hot Jupiters are  the most well-characterised planets as they are easy to detect and to follow-up with ground- and space-based telescopes.  Complementary to their bulk composition, information on their atmospheric properties and composition is also obtained based on atmospheric follow-up with e.g. Spitzer and HST,  with more data expected from JWST and ECHO. 
Gaining more and more information on exoplanet properties has opened new challenges that theoretical models now need to address. The anomalous radius observed for a significant fraction of hot Jupiters and related to the stellar incident flux is one of these challenges. 
Also, measuring the masses of Neptune-size  and smaller planets that receive high incident flux is now necessary to  probe the mass-radius-incident-flux relation for low-mass planets and to progress on this front.  Future transiting projects like  NGTS, CHEOPS and TESS, which will be devoted to super-Earth and Netpune mass planets, will provide in a near future an important contribution in this context. The detection of Earth-size planets remains a prominent observational goal and we are now getting the first harvest. But there is still a along way to go to perform   follow-up of  Earth-size planets and to characterise their atmospheres.  

Direct imaging of giant planets is starting to bear fruits and will become a major detection and characterisation technics in the coming decades with projects like Gemini Planet Imager (GPI), VLT-SPHERE and the coming next generation of Extremely Large Telescopes (ELT). As this technic directly probes planet atmospheres, it is on a better footing than indirect methods (secondary eclipse, transmission spectroscopy) to study in depth atmospheric properties (e.g chemical composition) and to infer global thermal properties of a planet (luminosity, surface temperature). It also completes the sample of planets detected by transit or radial velocity, mostly with young giant planets far from their parent star. This possibility to characterise planets just  after their formation process offers an exciting perspective to better understand planet formation and thermal evolution. 

\medskip

\textbf{ Acknowledgments.} 
This work is partly supported by the Royal Society award WM090065 and by the European Research Council under 
the  Union's Seventh Framework (FP7/2007-2013/ERC grant agreement no. 320478)


\bibliographystyle{ppvi.bst}
\bibliography{references}

\end{document}